%% file: main.tex
\documentclass[conference]{IEEEtran}

\IEEEoverridecommandlockouts
\DeclareRobustCommand{\IEEEauthorrefmark}[1]{\smash{\textsuperscript{\footnotesize #1}}}
\usepackage{enumitem} % Load enumitem package

\usepackage{cite}
\usepackage{amsmath,amssymb,amsfonts}
\usepackage{algorithmic}
\usepackage{graphicx, subcaption}
\usepackage{siunitx}
\usepackage{textcomp}
\RequirePackage[hyphens]{url}
\usepackage{xcolor}
\usepackage{siunitx}
\usepackage[colorinlistoftodos,prependcaption,textsize=tiny]{todonotes}
\usepackage[normalem]{ulem}
\usepackage{tikz}
\usetikzlibrary{arrows,arrows.meta,decorations.pathmorphing,backgrounds,positioning,fit,petri,calc,shapes.geometric,scopes,fit,shapes.geometric,math}

\usepackage[colorlinks=false,linkbordercolor=red,citebordercolor=green,filebordercolor=blue,urlbordercolor=white,pdfborder={2 1 0.5}]{hyperref}
\usepackage{orcidlink}
\usepackage{setspace}
\input{tikzFigsPicks.tex}

\newcommand\jksout{\bgroup\markoverwith{\textcolor{green!70!black}{\rule[0.5ex]{2pt}{0.4pt}}}\ULon}

\newcommand\richardsout{\bgroup\markoverwith{\textcolor{purple}{\rule[0.5ex]{2pt}{0.4pt}}}\ULon}
\newcommand\richardsUL{\bgroup\markoverwith{\textcolor{purple}{\rule[-0.2ex]{2pt}{0.4pt}}}\ULon}

\DeclareSIUnit\Tebibyte{TB}
\DeclareSIUnit\Gibibyte{GB}

\def\BibTeX{{\rm B\kern-.05em{\sc i\kern-.025em b}\kern-.08em
    T\kern-.1667em\lower.7ex\hbox{E}\kern-.125emX}}

\begin{document}

\title{The Artificial Scientist: in-transit Machine Learning of Plasma Simulations
}

\author{
\iffalse
\IEEEauthorblockN{Jeffrey Kelling}
\IEEEauthorblockA{\textit{Helmholtz-Zentrum Dresden-Rossendorf (HZDR)} \\
\textit{Chemnitz University of Technology} \\
Dresden, Germany \\
ORCID: 0000-0003-1761-2591} % j.kelling@hzdr.de
\and
\IEEEauthorblockN{Vicente Bolea} 
\IEEEauthorblockA{\textit{Kitware Inc.} \\ 
Clifton Park, NY, United States of America   \\
ORCID: 0000-0002-5382-093X} % vicente.bolea@gmail.com
\and
\IEEEauthorblockN{Michael Bussmann}
\IEEEauthorblockA{\textit{Helmholtz-Zentrum Dresden-Rossendorf (HZDR)} \\ 
\textit{Center for Advance Systems Understanding (CASUS)} \\
Dresden, Germany \\
ORCID: 0000-0002-8258-3881} % m.bussmann@hzdr.de
\and
\IEEEauthorblockN{Ankush Checkervarty}
\IEEEauthorblockA{\textit{Helmholtz-Zentrum Dresden-Rossendorf (HZDR)} \\
Dresden, Germany \\
ORCID: 0009-0004-4653-0004} % a.checkrvarty@hzdr.de
\and
\IEEEauthorblockN{Alexander Debus}
\IEEEauthorblockA{\textit{Helmholtz-Zentrum Dresden-Rossendorf (HZDR)} \\
Dresden, Germany \\
ORCID: 0000-0002-3844-3697} % a.debus@hzdr.de
\and
\IEEEauthorblockN{Jan Ebert}
\IEEEauthorblockA{\textit{Forschungszentrum Jülich} \\
J\"{u}lich, Germany  \\
ORCID: 0000-0001-7118-0481} % ja.ebert@fz-juelich.de
\and
\IEEEauthorblockN{Greg Eisenhauer}
\IEEEauthorblockA{\textit{Georgia Institute of Technology} \\
Atlanta, GA, United States of America  \\
ORCID: 0000-0002-2070-043X} % eisen@cc.gatech.edu
\and
\IEEEauthorblockN{Vineeth Gutta}
\IEEEauthorblockA{\textit{University of Delaware}\\
Newark, DE,  United States of America \\
ORCID: 0000-0002-1382-8128} % vineethg@udel.edu
\and
\IEEEauthorblockN{Stefan Kesselheim}
\IEEEauthorblockA{\textit{Forschungszentrum Jülich}\\
J\"{u}lich, Germany \\
ORCID: 0000-0003-0940-5752} % s.kesselheim@fz-juelich.de
\and
\IEEEauthorblockN{Scott Klasky} 
\IEEEauthorblockA{\textit{Oak Ridge National Laboratory}\\
Oak Ridge, TN, United States of America \\
ORCID: 0000-0003-3559-5772} % klasky@ornl.gov
\and
\IEEEauthorblockN{Richard Pausch}
\IEEEauthorblockA{\textit{Helmholtz-Zentrum Dresden-Rossendorf (HZDR)} \\
Dresden, Germany \\
ORCID: 0000-0001-7990-9564} % r.pausch@hzdr.de
\and 
\IEEEauthorblockN{Norbert Podhorszki}
\IEEEauthorblockA{\textit{Oak Ridge National Laboratory} \\
Oak Ridge, TN, United States of America \\
ORCID: 0000-0001-9647-542X} % pnb@ornl.gov
\and
\IEEEauthorblockN{Franz Pöschel}
\IEEEauthorblockA{\textit{Center for Advance Systems Understanding (CASUS)} \\
Görlitz, Germany \\
ORCID: 0000-0001-7042-5088} % f.poeschel@hzdr.de
\and
\IEEEauthorblockN{David Rogers} 
\IEEEauthorblockA{\textit{Oak Ridge National Laboratory} \\
Oak Ridge, TN, United States of America \\
ORCID: 0000-0002-5187-1768} % rogersdm@ornl.gov
\and
\IEEEauthorblockN{Jeyhun Rustamov}
\IEEEauthorblockA{\textit{Helmholtz-Zentrum Dresden-Rossendorf (HZDR)} \\
Dresden, Germany \\
ORCID: 0000-0002-4803-2461} % j.rustamov@hzdr.de
\and
\IEEEauthorblockN{Steve Schmerler}
\IEEEauthorblockA{\textit{Helmholtz-Zentrum Dresden-Rossendorf (HZDR)} \\
Dresden, Germany \\
ORCID: 0000-0003-1354-0578} % s.schmerler@hzdr.de
\and
\IEEEauthorblockN{Ulrich Schramm}
\IEEEauthorblockA{\textit{Helmholtz-Zentrum Dresden-Rossendorf (HZDR)} \\
Dresden, Germany \\
ORCID: 0000-0003-0390-7671} % u.schramm@hzdr.de
\and
\IEEEauthorblockN{Klaus Steiniger}
\IEEEauthorblockA{\textit{Helmholtz-Zentrum Dresden-Rossendorf (HZDR)} \\ 
\textit{Center for Advance Systems Understanding (CASUS)} \\
Dresden, Germany \\
ORCID: 0000-0001-8965-1149} % k.steiniger@hzdr.de
\and
\IEEEauthorblockN{René Widera}
\IEEEauthorblockA{\textit{Helmholtz-Zentrum Dresden-Rossendorf (HZDR)} \\ 
Dresden, Germany \\
ORCID: 0000-0003-1642-0459} % r.widera@hzdr.de
\and
\IEEEauthorblockN{Anna Willmann}
\IEEEauthorblockA{\textit{Helmholtz-Zentrum Dresden-Rossendorf (HZDR)} \\ 
Dresden, Germany \\
ORCID: 0009-0001-9992-6027} % a.willmann@hzdr.de
\and
\IEEEauthorblockN{Sunita Chandrasekaran} 
\IEEEauthorblockA{\textit{University of Delaware} \\
Newark, DE,  United States of America \\
ORCID: 0000-0002-3560-9428} % schandra@udel.edu
\else
\author{
\IEEEauthorblockN{
Jeffrey Kelling\IEEEauthorrefmark{1,2}, %\orcidlink{0000-0003-1761-2591} % j.kelling@hzdr.de
Vicente Bolea\IEEEauthorrefmark{3}, %\orcidlink{0000-0002-5382-093X} % vicente.bolea@gmail.com
Michael Bussmann\IEEEauthorrefmark{1,4}, %\orcidlink{0000-0002-8258-3881} % m.bussmann@hzdr.de
Ankush Checkervarty\IEEEauthorrefmark{1}, %\orcidlink{0009-0004-4653-0004} % a.checkrvarty@hzdr.de
Alexander Debus\IEEEauthorrefmark{1}, %\orcidlink{0000-0002-3844-3697} % a.debus@hzdr.de
\\
Jan Ebert\IEEEauthorrefmark{5}, %\orcidlink{0000-0001-7118-0481} % ja.ebert@fz-juelich.de
Greg Eisenhauer\IEEEauthorrefmark{6}, %\orcidlink{0000-0002-2070-043X} % eisen@cc.gatech.edu
Vineeth Gutta\IEEEauthorrefmark{7}, %\orcidlink{0000-0002-1382-8128} % vineethg@udel.edu
Stefan Kesselheim\IEEEauthorrefmark{5}, %\orcidlink{0000-0003-0940-5752} % s.kesselheim@fz-juelich.de
Scott Klasky\IEEEauthorrefmark{8}, %\orcidlink{0000-0003-3559-5772} % klasky@ornl.gov
Vedhas Pandit\IEEEauthorrefmark{1}, %\orcidlink{0000-0002-1983-8140} % v.pandit@hzdr.de
\\
Richard Pausch\IEEEauthorrefmark{1}, %\orcidlink{0000-0001-7990-9564} % r.pausch@hzdr.de
Norbert Podhorszki\IEEEauthorrefmark{8}, %\orcidlink{0000-0001-9647-542X} % pnb@ornl.gov
Franz Pöschel\IEEEauthorrefmark{4}, %\orcidlink{0000-0001-7042-5088} % f.poeschel@hzdr.de
David Rogers\IEEEauthorrefmark{8}, %\orcidlink{0000-0002-5187-1768} % rogersdm@ornl.gov
Jeyhun Rustamov\IEEEauthorrefmark{1}, %\orcidlink{0000-0002-4803-2461} % j.rustamov@hzdr.de
Steve Schmerler\IEEEauthorrefmark{1}, %\orcidlink{0000-0003-1354-0578} % s.schmerler@hzdr.de
\\
Ulrich Schramm\IEEEauthorrefmark{1}, %\orcidlink{0000-0003-0390-7671} % u.schramm@hzdr.de
Klaus Steiniger\IEEEauthorrefmark{1,4}, %\orcidlink{0000-0001-8965-1149} % k.steiniger@hzdr.de
René Widera\IEEEauthorrefmark{1}, %\orcidlink{0000-0003-1642-0459} % r.widera@hzdr.de
Anna Willmann\IEEEauthorrefmark{1}, %\orcidlink{0009-0001-9992-6027} % a.willmann@hzdr.de
Sunita Chandrasekaran\IEEEauthorrefmark{7} %\orcidlink{0000-0002-3560-9428} % schandra@udel.edu
}
\\
    \IEEEauthorblockA{
\IEEEauthorrefmark{1}\textit{Helmholtz-Zentrum Dresden-Rossendorf (HZDR)}, Dresden, Germany\\
\IEEEauthorrefmark{2}\textit{Chemnitz University of Technology}, Chemnitz, Germany\\
\IEEEauthorrefmark{3}\textit{Kitware Inc.}, Clifton Park, NY, United States of America\\
\IEEEauthorrefmark{4}\textit{Center for Advance Systems Understanding (CASUS)}, Görlitz, Germany\\
\IEEEauthorrefmark{5}\textit{Forschungszentrum Jülich}, Jülich, Germany\\
\IEEEauthorrefmark{6}\textit{Georgia Institute of Technology}, Atlanta, GA, United States of America\\
\IEEEauthorrefmark{7}\textit{University of Delaware} Newark, DE,  United States of America\\
\IEEEauthorrefmark{8}\textit{Oak Ridge National Laboratory} Oak Ridge, TN, United States of America
}
}
\fi
}

\maketitle
\begin{abstract}
Large-scale simulations or scientific experiments produce petabytes
of data per run. This poses massive challenges for I/O and storage when
scientific analysis workflows are run manually offline.
Unsupervised deep learning-based techniques to extract patterns and non-linear
relations from these large amounts of data provide a way to build scientific
understanding from raw data, reducing the need for manual pre-selection of
analysis steps, but require exascale compute and memory to process the full dataset available. 
In this paper, we demonstrate a heterogeneous streaming workflow in which plasma simulation data is streamed directly to a Machine Learning (ML) application training a model on the simulation data in-transit, completely circumventing the capacity-constrained filesystem bottleneck. This workflow employs openPMD to provide a high level interface to describe scientific data and also uses ADIOS2, to transfer volumes of data that exceed the capabilities of the filesystem.
We employ experience replay to avoid catastrophic forgetting in learning from this non-steady state process in a continual manner and adapt it to improve model convergence while
learning in-transit.
As a proof-of-concept, we approach the ill-posed inverse problem of predicting
particle dynamics from radiation in a particle-in-cell (PIConGPU) simulation of
the Kelvin-Helmholtz instability (KHI).
We detail hardware-software co-design challenges as we scale PIConGPU to full Frontier, the Top-1 system as of June 2024 Top500 list.

\end{abstract}


\section{Introduction}
HPC simulations capable of fully exploiting the compute power of supercomputers produce a wealth of high quality, complex data. 
Examples for such simulations include digital twins of the earth, whole device models of magnetic fusion reactors or simulations of advanced particle accelerators. The data rates and volumes of such simulations can be compared to instruments at large-scale research infrastructures such as tomography endstations at synchrotrons in photon science, high energy physics experiments, fusion research facilities or astro-particle observatories~\cite{bremer2023}.
Exploiting such a huge volume of data optimally often focuses on fast data selection including data filtering, triggering, anomaly and outlier detection or intelligent data reduction. In-situ inference using pre-trained machine learning (ML) models has led to a broad inclusion of such techniques across scientific communities~\cite{ZengZhuLam_2021DeepLearningDigitalHolograpReview,KocerKoBehler2022_NeuralNetworkPotentiaConciseOverviewMethods,DoppEberleHowardIrshadStreet2023_DataDrivenScienceMachineLearningMethods,BelisOdagiuAarres2024_MachineLearningAnomalyDetectioParticlePhysics}.

\begin{figure}[tbp]
    \centering   \includegraphics[width=8cm]{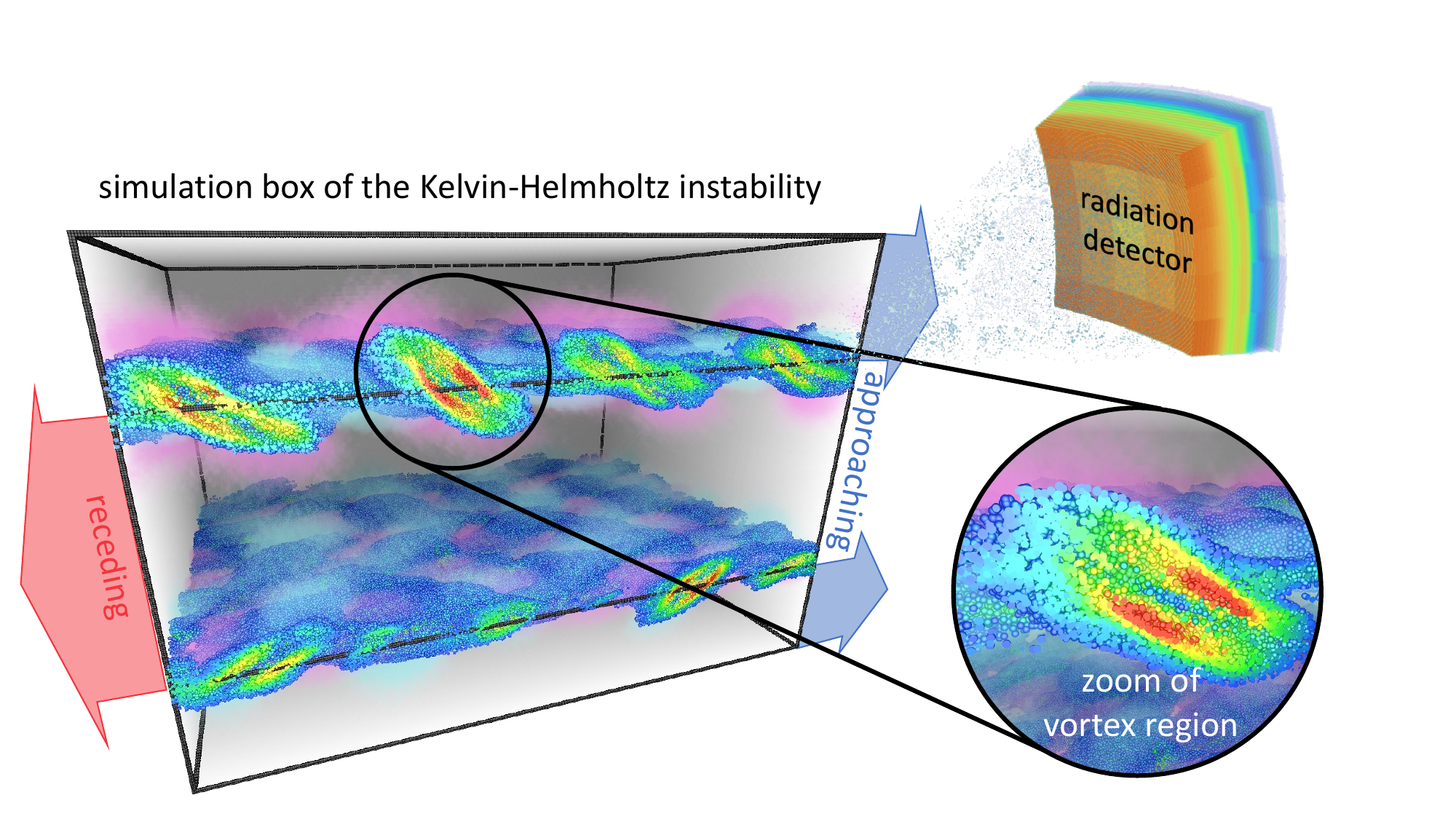}
    \caption{\small{3D rendering of the KHI simulated with PIConGPU~\cite{Bussmann2013} and rendered with ISAAC~\cite{Meyer2023}. 
    The middle graphic shows electrons as particles that radiate strongly, with the radiation intensity indicated from blue to red. 
    The initial plasma flow direction is depicted as arrows, blue marking propagation towards and red away from the radiation detector. The spectrally resolved radiation determined by the synthetic radiation detector is illustrated for a finite solid angle, showing radiation intensity per direction and frequency (indicated by depth).} 
    %The length of the box is $10.7 \frac{c}{\omega_\mathrm{pe}} \times 14.2 \frac{c}{\omega_\mathrm{pe}} \times 0.67 \frac{c}{\omega_\mathrm{pe}}$, with $\frac{c}{\omega_\mathrm{pe}}$ being the plasma skin depth.
    } 
    \label{fig:KHI_3D}
    \vspace{-0.5cm}
\end{figure}

Here we want to highlight an open challenge, namely the situation where a stream of
data produced by a source -- like a large-scale simulation or a high-resolution and high-repetition-rate experiment -- is of such a scale and quality that, even after simple data
reduction, storage for further offline analysis is impossible due to the physical
bandwidth or capacity constraints of the filesystem at hand. In such cases data
reduction already demands complex in-transit analysis. 

We illustrate this challenge with a particle-in-cell simulation of a
Kelvin-Helmholtz instability (KHI)~\cite{Sen1963} as shown in
Fig.~\ref{fig:KHI_3D} (more details in Section~\ref{sec:secKHI}). Instabilities
are one of the most fundamental, yet complex collective phenomena in plasma
physics. Understanding and controlling
them is critical across all disciplines of plasma physics from magnetic
and inertial fusion to space physics, plasma acceleration or astrophysics.
Instability growth rate, length scales and dynamics show a strong non-linear
dependence on the plasma properties.  
Reconstructing the complete phase space dynamics of a plasma from
observations would provide the most detailed view of the growth and
dynamics of instabilities, and thus of the most central plasma properties at high
temporal and spatial detail. 

Reconstructing the complete phase space dynamics of a plasma requires processing of a wealth of information. 
%One of the most demanding use cases in this class of problems is in-transit training of data-driven complex machine learning models at extreme data rates, where training the model requires the majority of data produced by the source. (RP: sonly mall changes for better readabilty, kept text for comparision)
One of the most challenging use cases in this class of problems is the
in-transit training of data-driven complex ML models at extreme data rates, where training the model requires most of the data produced by the source. 
The model being trained itself must be of sufficient complexity that it can, in principle, incorporate the complexity of the total volume of data produced by the source, while at no time a total view of the data is available. 

We note that a large number of advances in ML rely on iterative
training~\cite{Goodfellow-et-al-2016}, where the model repeatedly learns from
the input data over multiple epochs, requiring permanent storage of the
datasets. In the reference case studied in our work, this is not the case, as
data is produced on demand and discarded after being used for training. Due to
the size of data used for training and memory constraints, \emph{no total view} of the
training data is available at any given time. We thus have chosen a
continuous learning approach to address this challenge.
Although methods for continuous learning exist \cite{Gomes2019}, they have not had to consider the time constraints imposed by the speed of data generated from simulations~\cite{WangZhangSuZhu2024_CompreheSurveyContinuaLearningTheoryMethod}.
%{\color{blue}SC: do we have a citation for this sentence? is it Gomes2019, if yes it can be moved to the end of the line} {\color{red}RP: no Gomes2019 does not fit - Kelling please see todo---JK: Added a ref that is a pretty large review. They do not state that time contraints are never a relevant, but they also do not mention streaming performance or streaming from simulations} \todo[color=red]{Jeffrey: please cross check}
Thus, adapting the continuous learning techniques to handle such constraints is a critical challenge in the field of ML for high rate streaming of scientific data.

%The other challenge is to effectively tackle the memory hierarchy of heterogeneous compute systems; \textcolor{orange}{VG: doesn't flow well} they leads to huge differences in throughput \cite{Huebl_IO_scaling,Wan2022}. (old text for comparison)
%The other challenge is to effectively manage the memory hierarchy of heterogeneous computing systems, as it leads to enormous differences in throughput \cite{Huebl_IO_scaling,Wan2022}.\todo{@VG does it read better} 
%\jk{{\it how about:} 
%The other challenge is to effectively manage the memory hierarchy of heterogeneous computing systems, as it leads to enormous differences in throughput \cite{Huebl_IO_scaling,Wan2022}.\todo{@VG does it read better} 
A challenge to scaling is to effectively navigate the memory hierarchy of heterogeneous computing systems, as it exposes enormous differences in throughput \cite{Huebl_IO_scaling,Wan2022}. %}
While internal throughput in GPUs can easily reach the TB/s range on a single node, breaking down the throughput of massively parallel filesystems to the single node throughput reveals maximum performance of several tens of MB/s only. Each node often houses more than one GPU, hence available data storage and bandwidth demand data reduction, often done by discarding highly valuable data in practice. 

%We further note that the 
We further note that 
%this
mismatch in throughputs 
between memory hierarchies in modern HPC systems 
has been steadily growing, increasing the number of cases in which offline analysis is impossible and thus the urgency to find solutions for online analysis of data at high rates and volumes~\cite{Huebl_IO_scaling, pandit2019big}. 
We argue that in case of such extreme rates online analysis of streaming data is the method of choice~\cite{Poeschel2022, Meyer2023}, reducing the overall amount of data that needs to be stored while using computationally demanding algorithms to make optimum use of the data produced.
However, this means that a complete view of all data produced is never available at any time of the analysis workflow and there is a need
%and data production and analysis have 
to preserve the extreme data throughput during the whole analytics pipeline.

In our work we showcase a real-world example and present a general solution that implements all relevant parts of online in-situ training of a large-scale model on the Frontier exascale supercomputer at ORNL~\cite{olcf_frontier_url}. We argue that the building blocks of our solution are general enough to be adapted to almost any use case of the type discussed above with only minor adjustments. 
We furthermore emphasize that besides presenting a full workflow, the additional unique value of this work is to highlight the multi-parameter optimization problem of fitting the available compute and data transfer resources to achieve the best result in the shortest amount of time. 
We remark that our technique is especially interesting in the case of
irreversible data loss~\cite{Punch4nfdi}. It thus extends well beyond the
extreme case of large-scale simulations presented here. Examples include other high data rate 
sources, e.\,g.\, plasma or particle experiments with high-resolution and high-repetition-rate
detectors, which are more likely connected to smaller edge clusters.

\textbf{With a goal to automatically extract knowledge from large scale simulations, the paper makes the following contributions:}
\begin{itemize}

    \item \textbf{Computationally:} We present a workflow that integrates a particle-in-cell simulation of the Kelvin-Helmholtz instability (KHI), supported by openPMD-streaming\cite{Poeschel2022}, with a PyTorch~\cite{torchddp2_0}-based ML application \textit{(which we refer to as MLapp from here on)}. We regard this orchestration as the \textit{Artificial Scientist}. This orchestration, to the best of our knowledge, is the first of its kind. 
 \item \textbf{Scientifically:} We uncover correlations between emitted radiation and particle dynamics within the simulation in an unsupervised manner to create an inverse map from the observed radiation to the particle dynamics, and to identify the aspects of the particle dynamics that are relevant to the remotely observable radiation signatures of KHI.
 
 %  \item We demonstrate in-transit learning workflow at scale on Frontier (this is the first work of this kind, to the best of our knowledge). The orchestration of the different pieces of the workflow, i.\,e.\ KHI, streaming and ML application (MLapp), together is what we regard as - the \textit{Artificial Scientist}. 
   
   \item Using lessons learned from this orchestration and challenges faced, we provide input on what needs to be improved as we attempt to understand complex scientific data generated by instruments. 

\end{itemize}

%\section{Science case}
\section{The relativistic Kelvin-Helmholtz instability} \label{sec:secKHI}

The Kelvin-Helmholtz instability (KHI) is a well-known shear surface instability observed in plasmas~\cite{Sen1963}. 
It is driven by a self-amplifying cycle of small density or velocity fluctuations that lead to a growing magnetic field at the shear surface, which further amplifies the initial fluctuations (as depicted in Fig.~\ref{fig:KHI_3D}). 
This phenomenon occurs in astrophysical scenarios, including Saturn's magnetopause~\cite{delamere2011kelvin}, but also in fusion reactors and inertial confinement fusion (ICF)~\cite{casner2021recent, sadler2022faster}.
While the theoretical understanding and simulation capabilities of KHI \cite{Grismayer2013a, Alves2012} are well established, a direct measurement of the predicted density changes remains a real challenge. %, except in certain cases such as Saturn's magnetopause. 
Because of its well-understood dynamics, we chose KHI as a test case for our study to validate the results of our orchestrated workflow against existing knowledge.

A promising approach to the analysis of plasma dynamics is the study of the emitted radiation, which provides valuable information about the plasma properties and has been successfully applied in previous studies of the KHI~\cite{PauschPRE2017}. 
However, this correlation process typically requires extensive post-processing and analysis, often spanning several years.

Here, we simulate the KHI using the plasma simulation code PIConGPU \cite{Burau2010, Bussmann2013} which has been at the forefront of developing central technologies for GPU-accelerated plasma simulations, utilizing the alpaka library \cite{Zenker2016} for performance portability. 
It was the first particle-in-cell (PIC)~\cite{Hockney1988} code to run performant simulations on a variety of architectures in the world over many years~\cite{Meyer2023}.

% this could be removed since it just cleraly states wahat section III staes in more general terms
%The detection and characterization of disruptive instabilities such as the (sub-)relativistic KHI~\cite{Alves2012} is particularly important for fusion reactors, where such events can pose a serious safety risk. 
%Therefore, our work focuses on the automatic detection of correlations between plasma dynamics and emitted radiation during the KHI.

\section{Reconstructing the local phase space dynamics of the relativistic KHI from observable radiation \label{sec:reconstructing-local-phase-space-dynamics}}

\begin{figure}
 \DeclareRobustCommand\roundedBox{\resizebox{!}{.7em}{\tikz\draw[actor,modelDraw, rounded corners=.8ex] 
 rectangle (1.5em, 0.7em);}}
 \DeclareRobustCommand\sharpBox{\resizebox{!}{.7em}{\tikz \draw[modelDraw]
 rectangle (1.5em, 0.7em);}}
 \centering
 \figModelWorkflow
 \caption{\label{fig:modelTasks}%
 \small{
 Tasks that an ML model can be trained on, based on the simulation data
 in-transit:
 (a) solving the inverse problem of predicting particle dynamics from
 radiation signatures;
 (b) extracting features of particle dynamics and reconstruction thereof
 (representation learning/compression);
 (c) surrogate model of radiation emitted by complex particle dynamics.
Rounded boxes (\,\roundedBox\,)
indicate inputs/outputs, sharp boxes (\,\sharpBox\,) represent neural
networks. Same colors mark same data types/neural nets.
}}
\end{figure}

In this section, we will showcase the challenges faced and solutions employed to solve an extreme-scale inverse problem via a complex HPC workflow coupling a GPU-accelerated plasma simulation (KHI/PIConGPU) to a large-scale in-situ neural network for in-situ learning. This workflow as a whole, is a novel approach, to the best of our knowledge.  

%\jk{
Fig.~\ref{fig:modelTasks} depicts the interrelations between observational data (such as radiation) and simulation state (such as particles) as both inputs and outputs to data-driven models. Our objective of approximating the inverse mapping from observed radiation to particle dynamics can abstractly be described by the information flow depicted in Fig.~\ref{fig:modelTasks}(a). For the model to remain tractable a reduced representation of the complex particle dynamics is required, which is learned by an auto-encoding model Fig.~\ref{fig:modelTasks}(b). To observe the extent to which this representation captures relevant features of the particle dynamics, it is worthwhile to train a surrogate model of the computationally expensive, forward process of radiation emission (Fig.~\ref{fig:modelTasks}(c)). These three tasks integrate into our ML model, described later, also in Fig.~\ref{fig:network}.
%}

%Utilizing the alpaka library \cite{Zenker2016} it was the first plasma simulation code to run performant simulations across various hardware architectures using a single code base, powered by alpaka~\cite{Zenker2016}, frequently demonstrating excellent scaling and performance on the top ten high performance compute systems in the world over many years \cite{Meyer2023}.

%It has been observed early by us\todo[color=red]{here we might reveal us! SC: yes, I thought the same, I updated the sentence} and others that 

It was shown in 2013~\cite{Bussmann2013} that the computation of high quality, high resolution observational data such as radiation spectra can only be performed in-situ, as storing the ground truth data, the phase space trajectory data of each particle in the simulation at each time step, is impossible due to the aforementioned throughput hierarchy. As an example, scaling to even a moderate 25\,\% of the Frontier system, one would be faced with 1\,PB of data for every time step. With total time steps per simulation of the order of a few thousand time steps and time step durations of the order of 0.1 to \SI{1}{s}, we frequently observe data rates of 1 to 10\,PB/s for particle trajectory data and theoretically require total volumes on disk of the order of 10\,EB. This means the storage could get depleted for a simulation that is as short as \SI{100}s in duration.
%While coupling the calculation of derived observational data with the plasma simulation in a strongly coupled fashion on the GPU allowed calculating the spectra, eliminating the need to store trajectory data that could not be physically stored, this advance created a new challenge to use the spectral data to reconstruct plasma phase space information. Training an ML model for this task requires to connect the phase space data to the radiation data in order to allow for inversion, but it was clear that this phase space data could not be stored, which is a major bottleneck.
We thus end up with the conundrum to employ data-driven models for solving a large-scale ill-posed inverse problem without sufficient memory. % without data.

\subsection{A complex in-memory workflow to reconstruct local phase space dynamics from radiation \label{subs:complex-in-memory-workflow}}

A particular challenge for the reconstruction of local phase space dynamics is that their radiation signatures are highly convoluted, composite quantities that combine multiple facets of the ground truth to be restored. This observation guides our approach at learning from a live PIC simulation, as the posed problem is now diagnosed as ill-posed~\cite{kabanikhin2011inverse}.
In detail, the observed radiation spectrum is both spectrally and angularly resolved and the observed intensity at each wavelength and direction varies between time steps according to the changes of the phase space dynamics. The observed intensity in each direction and at each wavelength is a weighted sum of all contributions from all particles.

%It is clear that the only way to train a data-driven model for reconstructing the phase space from the radiation is by streaming the particle and radiation data from simulation directly into the model for training. At this point it must be pointed out that the observed radiation spectrum is both spectrally and angularly resolved and the observed intensity at each wavelength and direction varies between time steps according to the changes of the phase space dynamics. The observed intensity in each direction and at each wavelength is a weighted sum of all contributions from all particles, and as such a highly convoluted quantity. 
%Therefore, a solution that solely relies on independent algorithms each operating on a spatially localized region without communication will by the nature of the problem, not work.
%\todo[color=pink]{
%Kelling: please read this last sentence and determine wether it conflicts with the approach chosen
%---JK: 1. The paragraph does not describe the problem we are actually looking at, but this may be OK if it is supposed to give a bigger picture of longer term challenges. 2. The point "solely relies ..on a specially localized region [does not work]" is in conflict with the approach and the premise of our use of CL: if we look at the whole simulation volume at once, we are back to having only one train g sample per time step, if we learn from one simulation. This can only work in a steady-state system (in conflict with the abstract).}

\begin{figure}
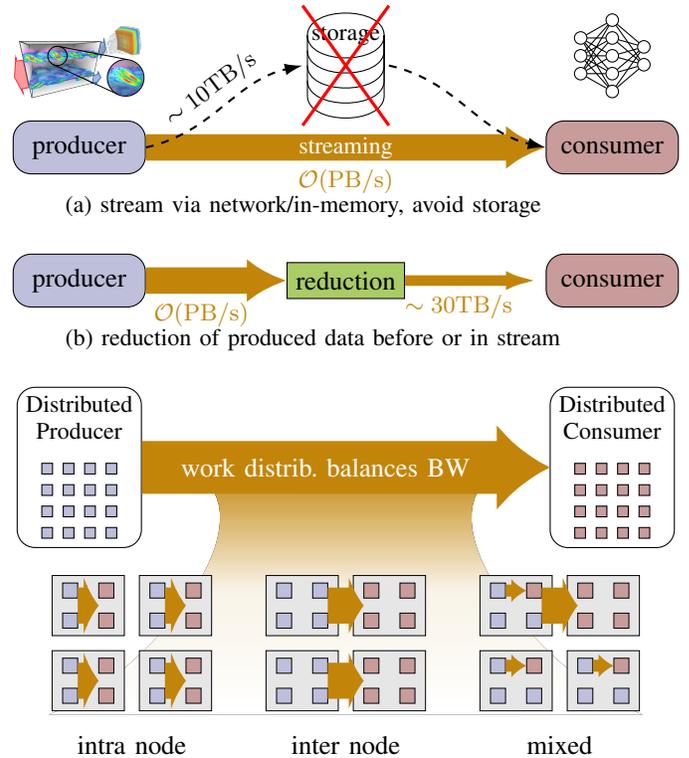

 \centering
 \figStreamSolutions
 \caption{\label{fig:streamingsolutions}%
\small{ Three aspects to streaming between loosely coupled producer and consumer: 
 (a) streaming without going through storage unlocks more bandwidth;
 (b) reducing simulation data close to the producer lowers bandwidth requirements;
 (c) to distributed producer and consumer, system topology presents
 communication paths with vastly different bandwidths which must be reconciled
 with the loosely-coupled application's communication requirements.}}
\end{figure}

Hence, here are the steps to be taken: 
\begin{itemize}
    \item Simulate a sufficiently large domain at a high spatial and temporal resolution to capture both the local plasma dynamics, as well as the collective contributions of different plasma regions to the total radiation spectrum.
    \item At each PIConGPU time step:
    \begin{itemize}
        \item Calculate the radiation spectrum in-situ as the sum of the contributions from all radiating particles in the simulation at each time step, with high angular and spectral resolution and high bandwidth.
        \item Collect the particle phase space data (namely the positions and momenta) and the complex amplitude and phase for each wavelength and direction of the radiation.
        \item Prepare the collected data for an ML model by finding suitable encodings  for spectral and phase space data.
        \item Stream the encoded data to an ML model and asynchronously train the model.
    \end{itemize}
    \item Stop the simulation after observing and covering a sufficient amount of relevant stages of plasma instability simulation.
%    observing the complete physical process under investigation.
    %collecting enough training data for the inversion and test the trained model %\todo[color=green]{SC:pls update the bullet based on the comment}\todo[color=yellow]{The qualification "for the inversion" is not true for non-steady-state problem and in slight contradiction to using continual learning (CL): Outside of a steady-state configurations are not repeated, hence running more time steps, does not produce more training data for a given time-step's phase-space state distribution. Only running a larger simulation of repeating cells does. If the problem were ergodic, experience replay would be dispensible if the auto-correlation times of the system are not too long (a limit that could probably be pushed towards infinity by adjusting the learning rate).}
\end{itemize}

\subsection{The technical challenge: Exascale data streaming and continual learning \label{subs:technical-challenge}}
An important design goal for the implementation of the above data processing scheme is to \textbf{write no simulation data} to the filesystem during the entire workflow. File I/O can certainly be initiated when desired, but is not necessary for the workflow.
%It should be pointed out that at no time in this workflow is any data written to the filesystem. 

\textbf{In-situ} approaches have been previously used to compute scientifically relevant analyses through plugins within the same application context~\cite{Bussmann2013}. This is no longer possible for the envisioned ML analysis, as the vastly incompatible software stacks for the C++ simulation code PIConGPU and the PyTorch-based MLapp, as well as their different scaling properties inhibit such an integration into a \textbf{strongly-coupled} monolithic application framework. Hence, the simulation and the MLapp are modularized into two separate applications that communicate through a standardized data interface -- a pattern that we refer to as \textbf{loose coupling}, since the standardized data interface allows to plug together a data processing pipeline more flexibly, i.e.\ loosely than before. When the implementation for such workflows avoids writing intermediate data to disk, we refer to them as \textbf{in-memory}. In-memory workflows include such setups where data is not moved, staying within one application's memory \textbf{(in-situ)}, and alternatively setups that move data from one into another application's memory \textbf{(in-transit)} (Fig.~\ref{fig:streamingsolutions}(a)). The latter becomes necessary when transitioning from file-based workflows to in-memory workflows without sacrificing flexibility in the data exchange pattern, as well as in response to codes that scale differently.

The whole simulation, all data produced by the simulation and ingested by the model, and the whole model, all reside in-memory and data is distributed and optionally reduced in-transit (Fig.~\ref{fig:streamingsolutions}(b)).

The physical workflow outlined in \ref{subs:complex-in-memory-workflow} can be directly translated into a technical setup that is the main subject of this work:

\begin{itemize}
    \item Start a scalable, GPU-accelerated PIConGPU simulation of sufficient size and resolution, utilizing a large %\todo[color=orange]{only <100 nodes could be used.SC: Yes, replaced signifiant with large as 100 is still a large set of nodes, not significant;-)}  
    subset of nodes on Frontier~\cite{frontier_docs}.
    \item Schedule PyTorch and N/RCCL (i.\,e.\, NVIDIA and ROCm collective communication library) alongside the PIConGPU simulation.
    \item Optimize scheduling for data transport and locality, physical system size and resolution, simulation and training time, total time of solution and use of resources (options in Fig.~\ref{fig:streamingsolutions}(c)).
    \item At each simulation time step:
    \begin{itemize}
        \item Stream particle momenta, particle positions and the spectral data to a large-scale ML model,
        \item Transform the phase space and spectral data from its simulation format to an optimum format for the ML mode and
        \item Train the model concurrently.
    \end{itemize}
    \item Repeat the workflow for a sufficient number of time steps to cover all relevant stages of the plasma instability simulation with respect to the dynamical evolution and amount of training data needed.
\end{itemize}

For supporting an online workflow which stores no intermediate data such as phase space trajectory and radiation data to disk, the model must be continuously trained online from the series of time step. This requires the model to be trained by subsequent snapshots of the evolution of plasma and radiation data. For this, continual learning is the methodology of choice as outlined in Section~\ref{machine_learning}.

We strongly believe that a large class of inversion problems for complex, large-scale systems that produce data streams face similar challenges, and that our solution presented here for the first time introduces all the relevant building blocks and technologies to address these challenges.

\section{Methodology}
\subsection{Particle-in-cell simulation and its setup}

\iffalse 
\textit{2-3 paragraphs = 0.75 column} \textbf{Richard, Klaus}

\begin{itemize}
    \item Explain that we use the particle-in-cell code PIConGPU \cite{Bussmann2013} for simulating the Kelvin-Helmholtz instability
    \item Briefly explain how the PIC cycle allows a PIC code to iteratively model plasma dynamics
    \item State that PIConGPU has demonstrated excellent weak and strong scaling on various clusters.
    \item Explain specifically what methods and libraries we use to reach such a good scalability. 
    \item State that PIConGPU supports a number of in-situ processing and analysis methods, from which we use the openPMD-streaming (see \ref{openPMD_streaming}) and the far-field radiation plugin for this study.
    \item Briefly describe how the far-field radiation plugin works.
    \item Explain the computational need to compute radiation in-situ and link to the need of ML approach.
    \item \textit{Plot: PIC-cycle and software stack}
\end{itemize}
\fi 

To simulate KHI, we employ the fully kinetic 3D3V relativistic particle-in-cell (PIC) method \cite{Hockney1988} using the open-source code PIConGPU \cite{Bussmann2013, Burau2010}.
%hosted and developed by Helmholtz-Zentrum Dresden-Rossendorf (HZDR)
%This approach linearizes the solution of Maxwell’s equations on a grid while representing particles as macro-particles that move freely within the simulation volume. 
The PIC method iteratively models plasma dynamics by advancing particle positions and velocities in response to electromagnetic forces, while the electromagnetic fields are updated based on the electric current caused by the particle motion. 
%In this simulation, the Boris pusher \cite{Boris1970} is used to update the particle dynamics, the Esirkepov scheme \cite{Esirkepov2001} for current deposition, and the Yee algorithm \cite{Yee1966} to update the electric and magnetic fields. 

For modeling the KHI with PIConGPU, the smallest volume we simulate is $192\times256\times12$ cells on 16 AMD MI250X GPUs using cubic cells of $\Delta x = 93.5 \, \mathrm{\mu m}$ and a time step duration of $\Delta t = 17.9 \mathrm{fs}$ at a particle density of $n_0=10^{25} \, \mathrm{m}^{-3}$. 
We initialize two counterpropagating plasma streams (as shown in Fig.~\ref{fig:KHI_3D}) with a normalized velocity of $\beta = \frac{v}{c} = 0.2$ and initialize 9 particles per cell.  
%\todo[color=orange]{Does it make sense to just describe the PIC setup here?}
%\todo[color=green]{I think so, perhaps the title can be Particle-in-cell simulations and setup?}

\begin{figure}[tbp]
  \centering
%  \begin{subfigure}[c]{0.45\textwidth}
%    \includegraphics[width=\textwidth]{plots/summit_weak_scaling.png}
%    \caption{}
%    \label{fig::SummitScaling}
%  \end{subfigure}
%  \hfill
%  \begin{subfigure}[c]{0.45\textwidth}
    \includegraphics[width=5.5cm]{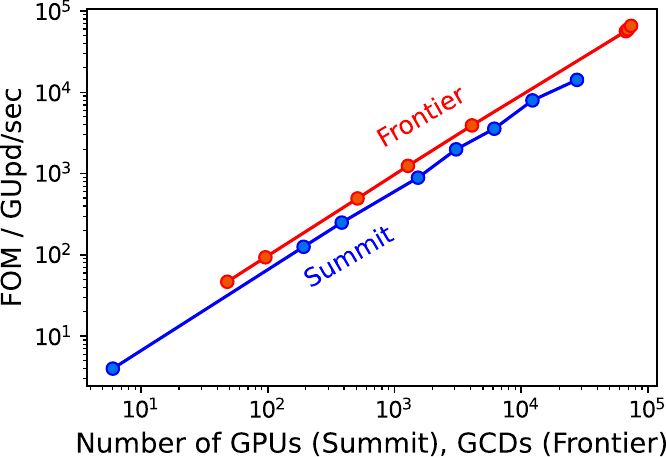}
    %\caption{}
   
%  \end{subfigure}
  \caption{
%    (a) Weak scaling of PIConGPU from 162 (27 nodes) to \num{27,600} NVIDIA V100 GPUs (4,600 nodes) on Summit (blue line). PIConGPU achieves a weak scaling efficiency $>95\,\%$. PIConGPU achieves an average FOM for the largest run of 14.7\,TeraUpdates/s.
 %   (b) 
 \small{FOM scaling of PIConGPU from 24 GPUs (6 nodes) to \num{36864} GPUs (\num{9216} nodes) on Frontier (red line) using a general test case~\cite{TWAECtestcase}.
    PIConGPU achieves an average FOM for the largest run of 65.3\,TeraUpdates/s vs 14.7\,TeraUpdates/s on Summit.
  }}
    \label{fig::frontierScaling}
    \vspace{-0.2in}
\end{figure}

To prove that PIConGPU can scale on a large system for general plasma physics cases, we use a more challenging test case~\cite{TWAECtestcase} than the KHI as a scaling benchmark, with a higher particle-per-cell ratio. 
Fig.~\ref{fig::frontierScaling} shows scaling runs %of PIConGPU %TWEAC on %large full scale systems such as 
on ORNL's Summit and Frontier. 

%PIConGPU has proven scalability on large-scale systems, such as ORNL's Summit and recently
%, as one of the OLCF Frontier Center for Accelerated Application Readiness teams, 
%on the full ORNL Frontier system.
%The code has been awarded with 1M node hours for an INCITE allocation on Frontier in 2023. \todo[color=orange]{should we mention that here? SC: it is better we don't, I commented it out}
%Recent weak scaling results of compute performance on Summit are shown in Fig.~\ref{fig::SummitScaling}.

On Frontier with a total of \num{36864} AMD MI250X GPUs across \num{9216} nodes~\cite{frontier_docs}, a scaling of PIConGPU's Figure of Merit (FOM), i.\,e.\ the weighted sum of the total number of particle updates per second (90\,\%) and the number of cell updates per second (10\,\%), is demonstrated as shown in the weak scaling results. %, Fig.~\ref{fig::frontierScaling}. 
For the results presented in this paper, the simulation involved \num{2.7e13} macroparticles in $10^{12}$ grid cells,
compared to our base FOM run on Summit in 2019, where only \num{{}e13} macroparticles in \num{4e11} cells could be used~\cite{budiardja2023ready}.

On Frontier, one thousand time-steps completed in a \textit{mere} 6.5\,minutes. % \textcolor{red}{Should we also write a line here about the peak/sustained power consumption of PIConGPU on Frontier}
This is due to several factors. Spatial domain decomposition distributes computational domains across GPUs and computers, exploiting data locality and enabling parallelization. 
Efficient data structures, such as supercells in PIConGPU \cite{Hoenig2010}, optimize data access patterns within these domains, while asynchronous communication strategies between compute nodes minimize communication overhead when exchanging information between different domains. 
In addition, usage of the libraries PMacc~\cite{libPMacc} and alpaka
\cite{Zenker2016} abstract data management and hardware-specific optimizations,
enabling portability across architectures.

PIConGPU uses openPMD \cite{openPMDapi, huebl2015openpmd} to output data.
By using openPMD, we stream particle data in-transit via ADIOS2~\cite{godoy2020adios} (\textit{see Section~\ref{openPMD_streaming}}) directly to the ML framework, thus avoiding filesystem limitations \cite{Huebl_IO_scaling}.
In addition, PIConGPU supports strongly-coupled in-situ processing and analysis methods, among which we use the far-field radiation plugin for our study. 
%\sunita{SC: do we need the next few lines? "In contrast.... incoherent radiation" can we shorten these few lines to ONE line that leads to the last line on computational cost?}
%In contrast to the intrinsic field solver, the far-field radiation plugin \cite{Pausch2014} calculates radiation emissions based on the Liénard-Wiechert potential approach \cite{Jackson1998}, thus resolving radiation spectrally and directionally. 
%This overcomes a common limitation of the discretized field solver of the PIC algorithm, since the plugin can resolve much higher frequencies and directly predict observable spectra and allows correctly quantifying coherent and incoherent radiation \cite{Pausch2018NIMA}.
In contrast to the intrinsic field solver, the far-field radiation plugin \cite{Pausch2014} calculates radiation emissions using the Liénard-Wiechert potential approach \cite{Jackson1998}, thus overcoming common limitation of the PIC algorithm, as it resolves much higher frequencies, directly predict observable spectra, and correctly quantify coherent and incoherent radiation \cite{Pausch2018NIMA}.
However, it comes at a high computational cost, scaling with the number of particles, frequencies, and directions, easily exceeding the computational limits of the underlying particle-in-cell simulation \cite{Pausch2014ipac}.

Since the computational cost of the far-field radiation plugin can be significant, forward calculations can become prohibitively expensive for a wide range of simulation parameters. 
Furthermore, inversion from observed radiation back to particle dynamics remains a challenging task, requiring ML-based innovative approaches.
Thus, we implemented a novel ML approach (\textit{see Section \ref{machine_learning}}) that was trained on both particle distributions and their associated radiation using streamed data from PIConGPU. 
After training, the ML model should generate particle distributions based on the input spectra, providing a potential solution to the inversion problem and improving our ability to interpret plasma dynamics based on observable radiation.

\subsection{Data streaming via openPMD} \label{openPMD_streaming}

Considering that both PIConGPU as well as the MLapp rely significantly on very different software stacks, launching both parts of the setup in the same application context is not feasible. As detailed previously in \ref{subs:technical-challenge}, we follow a loose-coupling rather than a tight-coupling approach \cite{Poeschel2022}, launching both applications separately.
Output created by PIConGPU must be moved across the application boundary to the MLapp, without impacting the scalability of the setup through use of limited resources such as the parallel filesystem. The solution pursued in this paper is an \textbf{in-transit} workflow that keeps data in memory at all times and writes no intermediate data to disk.

The in-transit I/O solution is put in place by  openPMD\cite{openPMDapi} implemented within PIConGPU and the MLapp. openPMD~\cite{huebl2015openpmd} is a data standard for particle-mesh data built for F.A.I.R.\ scientific I/O in HPC software that needs to scale, and it is employed in a diverse ecosystem of simulation, analysis and visualization codes\cite{openPMDprojects}. {In turn, the I/O building block for the orchestration of the \textit{Artificial Scientist}is modular and generically reusable for other setups of this kind.} 

\begin{figure}[tbp]
    \centering
    \includegraphics[width=3in]{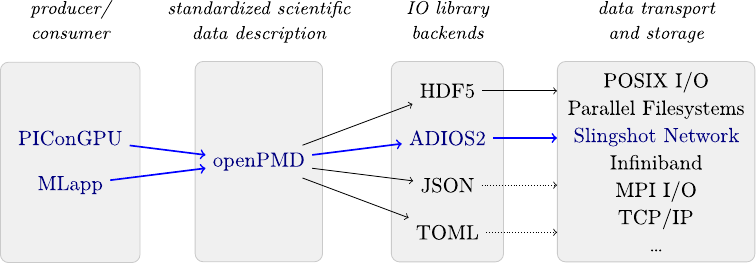}
    \caption{\small High-level overview of the software stack for data exchange between PIConGPU and the MLapp.}
    \label{fig:software-stack}
    \vspace{-0.2in}
\end{figure}

As an open \emph{data} format, openPMD can be implemented in various \emph{file} formats such as JSON, TOML, HDF5 and ADIOS2, all supported by the reference implementation openPMD-api, of which we further consider ADIOS2 in this study due to its support for in-transit I/O in HPC settings, as shown in Fig.~\ref{fig:software-stack},
% Using this software stack as well as the flexible choice of backends and I/O engines provided by the openPMD-api and ADIOS2, PIConGPU has a scalable I/O system, quickly adaptable to specific requirements of a software setup and of a compute system. 
    This flexible choice of backends for scalable and quickly adaptable I/O solutions gains outstanding relevance on exascale like Frontier where even modern solutions for classical I/O via parallel filesystems, e.\,g.\ the provisioned Orion filesystem, cannot be expected to scale up to full system size as a base for a loosely coupled data analysis pipeline.

The in-transit data strategy is built upon the Sustainable Staging Transport (SST)~\cite{Gainaru2022, eisenhauer2024streamingdatahpcworkflows} data engine of ADIOS2 that connects one parallel data producer code to an arbitrary number of parallel data consumer codes, thereby opening connections only between processes that actually share data. 
For adapting to different system architectures, the SST engine implements different network transport technologies (\emph{data planes}), including TCP (non-scalable fallback), libfabric, ucx and the \texttt{MPI\_Open\_port()} API of MPI.
For this benchmark, we consider the low-level libfabric data plane on top of the system's CXI provider \cite{libfabric_github} which allows, yet also requires more manual finetuning, as well the MPI data plane which leverages the MPI implementation's performance tuning.
%While the MPI data plane indeed works well on Frontier, the rather seldom-used port API of MPI tends to not be supported on many other HPC systems, and was unstable during the conception of this study. For these reasons, we additionally added support for Frontier's Slingshot network via libfabric's CXI provider \cite{libfabric_github} to ADIOS2, which now gives us the chance to compare performance numbers between different implementations. With the more low-level libfabric, we can circumvent further performance-impacting requirements by the MPI data plane, such as the use of threaded MPI or required writer handshake interaction for remote read accesses.

The use of HPC interconnects such as the Slingshot network on Frontier or Infiniband on e.\,g.\ Summit is an important cornerstone in the general applicability of this I/O building block, since this I/O approach thus retains the flexibility of setups based on the parallel filesystem. In particular, there is no lock-in to a specific pattern for data exchange, compared to approaches based on e.\,g.\ local SSDs or shared memory. While local data exchange within a node remains preferred for scalability, this I/O approach naturally extends towards patterns such as staging within a neighborhood of nodes (for scheduling reasons or for implicit load balancing via streaming) or a fan-in pattern (for data reduction purposes), both of which are potential directions to pursue. 

 % We measure the \emph{perceived throughput} based on the time used in the data reader for loading the particles. Since it also includes communication overhead, this perceived throughput is a lower bound on the actual throughput achieved. 

Before applying this I/O solution to the MLapp, 
%final PIConGPU+TC pipeline, 
we demonstrate its scalability to the full system using a synthetic benchmark that runs PIConGPU's KHI scenario full-scale and streams its particle data into a synthetic no-op consumer that performs no computation beside measuring the performance of this I/O operation and only discards received data. 
Employing the no-op consumer gives us a testbed for full-system scaling runs of a particle data stream fed by PIConGPU, helping us identify and eliminate scaling issues before applying the full PIConGPU+MLapp pipeline.
In that context, the major challenge is to scale to a large number of parallel instances, while the amount of data streamed per instance comes as a secondary scaling target. 

%\textcolor{red}{We need to cross cite figure 6 within the narrative. What is a no-op consumer in the figure? Hmm "The synthetic consumer
%runs with one instance per node on the CPU side" - not following this? Should we color code GPU CPU instances? PIC can be violet. But some red and beige color for AMD GPU and CPU perhaps? }
%\commentfranz{I followed the color distribution from figures 3 and 9 which have the producer in blue and the MLapp in red. For the consumer, I picked some other color since it is only a dummy and not the MLapp. The graphic is supposed to illustrate the dataflow between applications, and I don't want to make the GPU/CPU setup the focus here, hence the text mentions that PIConGPU uses the GPUs, but the graphic is about something else.}

% instead use \ref{fig:streamingsolutions}(c) (bottom left).
%\begin{figure}
%    \centering
%    \includegraphics[width=\linewidth]{pipe_setup_no_filesystem_cropped.pdf}
 %   \caption{\itshape Schematic setup for a full-scale streaming I/O benchmark on OLCF Frontier. Each instance of PIConGPU uses one GCD of AMD MI250X GPUs for compute acceleration, leading to eight instances per compute node. The data created by these eight instances is collected into one instance of the synthetic no-op consumer that only discards the received data.
 %   \label{fig:setup-benchmark-frontier}
 %   }
 %   %\vspace{-0.5cm}
%\end{figure}

The PIConGPU KHI simulation is run from half (4096 compute nodes) to full-scale (9126 compute nodes) of Frontier, producing \SI{5.86}{\Gibibyte} of particle data per compute node and time step.
Fig.~\ref{fig:streamingsolutions}(c) (bottom left) %\ref{fig:setup-benchmark-frontier} 
shows the processes running within each compute node and the data flow between them. At each scaling run (weak scaling), five time steps are sent from PIConGPU to the no-op consumer, which then measures the time needed for loading the data. The parallel throughput is calculated based on this measured time and the global data size, i.\,e.\ the previously mentioned \SI{5.86}{\Gibibyte} per node multiplied with the number of nodes. Although including communication overhead, this throughput value has been shown in \cite{eisenhauer2024streamingdatahpcworkflows} to be a close approximation for the real throughput in this kind of setup.

\begin{figure}[tbp]
\centering
\includegraphics[width=\linewidth]{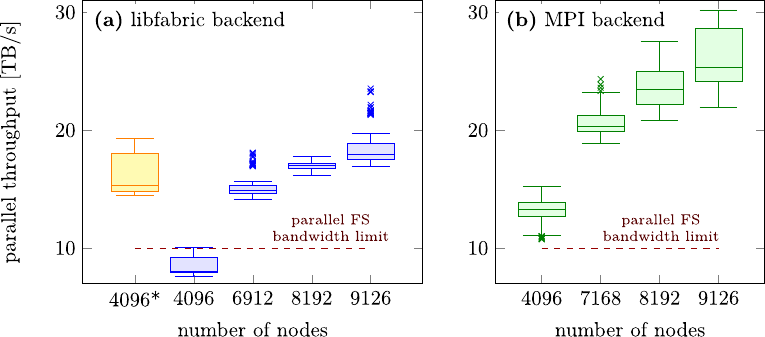}
\caption{\small{Parallel total throughput for streaming at full scale on Frontier using a synthetic benchmark built on PIConGPU KHI. An obvious outlier result was removed for the libfabric backend at 8192 nodes.  (a)~Using the libfabric data plane and the CXI provider to access the Slingshot network at a lower level. (b)~Using the MPI data plane based on \texttt{MPI\_Open\_port()}.  
%There are two runs at 4096 nodes; the first of them (marked with an asterisk) enqueued all read operations to the network at once, while the others enqueued them in batches and waited on them. \\
The parallel throughput reaches around 20TB/s for libfabric, and from 20 - 30 TB/s for MPI. }
}
\label{fig:synthetic-benchmark}
\vspace{-1cm}
\end{figure}

The benchmarks in Fig.~\ref{fig:synthetic-benchmark} demonstrate the feasibility of this I/O strategy for full-scale workflows.  Depicted are boxplots of all single measurements. Most notably, we observe a maximum parallel throughput of 20 - 30 TB/s which compares outstandingly against the 10 TB/s bandwidth of the parallel Orion filesystem, a scaling limit which we circumvent and exceed by not using the filesystem for intermediate data. Furthermore, these results can also compare well against the 35 TB/s  aggregate write bandwidth of the SSDs installed locally in the compute nodes~\cite{frontier_pfs_bandwidths}.

For the libfabric-based benchmarks, we initially used an implementation that enqueued all read operations to the network at once and waited for replies (labeled $4096 *$ in Fig.~\ref{fig:synthetic-benchmark}). While these results signify the best achieved per-node throughput of 3.5 - 4.7 GB/s at 4096 compute nodes \textit{(a total of 14.1 - 18.8\;TB/s)}, it turned out that this strategy did not scale to the full system. Thus, we employed an alternative that enqueued operations in batches of 10 operations. While this turned out to scale to the full system, it came at a notable cost in performance, visible in a per-node throughput of 1.9 - 2.6 GB/s at 9126 compute nodes \textit{(a total of 16.5 - 23.0\;TB/s)}. 
Conversely, the MPI data plane yields a per-node throughput from  2.6 - 3.7 GB/s at 4096 compute nodes \textit{(a total of 10.5 - 14.9\;TB/s)} to  2.4 - 3.3 GB/s at 9126 nodes \textit{(a total of 21.4 - 29.5\;TB/s)}, \textbf{the best achieved system-wide parallel throughput} from this experiment. 

We conclude that the MPI data plane brings default good performance, while the libfabric data plane's lower-level control can bring performance improvements but at the cost of necessary fine-tuning beyond this study's scope.
Since only a single instance of the reader code was running per node, its throughput (\num{1.9} - \SI{4.7}{GB/s}, across all cases) compares against the max possible throughput of a single HPE Slingshot NIC at $\SI{25}{GB/s}$. This implies that further speedup can be achieved by parallelizing the reader in the actual PIConGPU+MLapp pipeline.
The benchmarks suggest that the MPI data plane be used, which can be explained by leveraging system-specific performance tuning in the MPI implementation. All regular measurements range between \SI{1.2}{\second} - \SI{3.2}{\second}.

\subsection{Machine learning} \label{machine_learning}
The task of in-transit ML is to extract patterns from the live observations and thereby learn correlations between variables and reduced representations. 
Here, we focus on predicting particle dynamics from radiation spectra, which constitutes an ill-posed problem -- since any given radiation spectrum can potentially be produced by various types of dynamics~\cite{kabanikhin2011inverse}. 
When learning in-transit from a simulation, this problem is reduced from an intractable general case to the question of: which type of local particle dynamics under the given physical setup, i.\,e.\ the relativistic KHI, gave rise to an observed radiation spectrum observed in one direction, which, however, remains an ill-posed problem~\cite{kabanikhin2011inverse}. 
Many ML techniques, especially feed-forward networks, tend to predict the mean of possible inversions, even if this itself is not a valid solution~\cite{GalGhahra2016_DropoBayesApproRepreModelUncer}.
% In architectures such as conditional variational auto encoders (cVAEs) or conditional generative adversarial networks (cGANs) this is a common type of mode-collapse.
One class of techniques specifically designed to learn by sampling from conditional multi-modal distributions, such as those that occur as solutions to our ill-posed problem, are flow-based models~\cite{DinhSohl-DBengio2016_DensityEstimatiUsingRealNvp,LuHuang2020_StrucOutpuLearnCondiGenerFlows} and invertible neural networks (INNs)~\cite{ArdizzKruseWirkerRahnerKothe2018_AnalyInverProblInverNeuraNetwo}. 
These learn an invertible transformation from a chosen posterior distribution (often a normal distribution $\mathcal{N}$) and a prior described by the training data (in our case the distribution of observed particle dynamics $\mathcal{D}$) and vice versa. 
One specific difference between them is that INNs also learn the deterministic forward function defining the inverse problem -- in our case, the mapping between samples of particle dynamics $\mathcal{D}$ and radiation spectrum $I$. 
Put differently, INNs also learn to predict the condition $I$ for a given prior $D$.

Particle configurations in phase space of the plasma dynamics simulated here can be very complex, such that a small sample of particles cannot adequately represent the dynamics. 
For INNs and flow-based models this poses a challenge, because the change-of-variables approach~\cite{TabakVanden2010_DensiEstimDualAscenLogLikel, TabakTurner2013_FamilNonpaDensiEstimAlgor} dictates that the information volume, i.\,e.\ the number of activations at the input and output of each coupling block, must be constant throughout the network, which results in a very large network when working with a prior distribution of dimensionality on the order of $10^5$.
In this setting, the number of network parameters can be kept in check by using affine flows~\cite{DinhKruegeBengio2015_NiceNonLinearIndependComponenEstimati,DinhSohl-DBengio2016_DensityEstimatiUsingRealNvp} with convolutional sub-networks designed to be invariant to transposition of particles in the input vector to be efficiently trainable.

%The usual way in which problem is simplified, which we also explored is to 
When generating point cloud (PC), flow-based models are typically tasked with sampling them point-by-point, i.e. the model learns a probability density describing the particle density in phase space. The main drawback of this method is that the model has to learn the many-modal distributions which can occur in the KHI phase space on a single-particle basis and can no longer generate multiple particle distribution for a single condition, i.\,e.\, we cannot expect it to distinctly produce all PCs solving an ill-posed problem ~\cite{kabanikhin2011inverse}. 
Another issue is that, unless we use an encoder (conditioning network) producing very small latent vectors of around six dimensions for the radiation, the dimension of the condition $\mathcal{I}$ is much larger than the dimension of the prior while the network has to be sampled $\mathcal{O}(10^3)$ times to retrieve a useful particle distribution.
%The network is then also effectively tasked with learning the conditional probability distributions of the particles %and struggles to learn the many-modal distributions which can occur in the KHI phase space.

To overcome the above inefficiencies, we opted to 1. view the particle configuration as a point cloud (PC), and 2. reduce the dimensionality of the prior $\mathcal{D}$ by tasking an outer network to learn latent representations of the phase space that are also invariant with respect to transposition of particles. 
This outer network can, in principle, be an autoencoder (AE) whose latent representation $z$ is passed to the INN as samples of the prior distribution $Z$. 
In practice, the INN will not generate exact latent vectors $z'$ on its backward pass, but variations, which will cause the AE's decoder to fail at reconstruction. 
Therefore, we employ a variational auto encoder (VAE) instead, which, albeit harder to train, intrinsically trains its decoder to be robust against the variations produced by the INN.

\begin{figure}[tb]
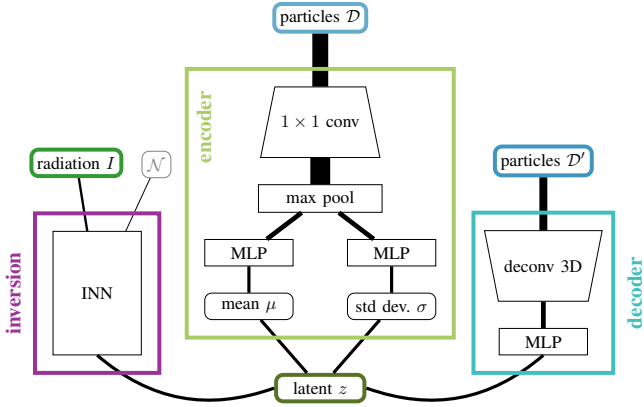

\centering
 \figModelIOOutBW
 \caption{\small{Block-wise architecture of the ML model encompassing all three tasks
 summarized in Fig.~\ref{fig:modelTasks} to
 allow in-transit training and regularize the pivotal latent space. Colored
 boxes correspond to the elements in Fig.~\ref{fig:modelTasks}. The black lines
 indicate information flow between blocks with their thicknesses illustrating
 the relative amounts of activations communicated. The combination of the encoder and decoder blocks constitute the VAE portion of the model.}}
\label{fig:network}
\vspace{-0.5cm}
\end{figure}

%{\color{blue}SC: need brighter colors, green is barely visible :-)can we shade each of those dotted boxes with earthy soft colors of green blue and red? \textit{--- e.\,g.\ \texttt{green!60!brown}?} Secondly, in the boxes for $\mu$  $\sigma$ and $z$, can we update it, similar to particles and radiation boxes where there is the symbol + what it means? so $z$ would be latent vector, mean $\mu$ standard deviation $\sigma$ \textit{--- JK: Yes, but "standard deviation" is very long, need to make the bock have two lines, then}} %SC - yes the boxes are good now, JK. 
%Fig.~\ref{fig:network} shows a block diagram of the full architecture, which decomposes into 
As depicted in  Fig.~\ref{fig:network}, there are three main blocks of the network architecture:
\begin{itemize}
\vspace{-0.03in}
    \item The encoder (light green) follows an architecture proposed in~\cite{KongRajakShaker2023_GenerModel3dPointCloud}, which is a simplified version of PointNet~\cite{QiSuMoGuibas2017_PointnetDeepLearningPointSets3d}. 
    6-dimensional $\num{3e4}$ vectors corresponding to the positions and momenta of the $\num{3e4}$ particles are fed to 
    $1\times1$ convolutions separately to extract 608 features per particle (channels: $6\to16\to32\to64\to128\to256\to608$), followed by a max-pooling to obtain a transposition-invariant feature set. These 608 features are then processed by two multi-layer perceptrons (MLP), each with $608\to544$ hidden features, to obtain predictions for the mean $\mu$ and standard deviation $\sigma$ to sample the 544-dimensional latent vector $z$.
    \item The decoder (cyan) uses a single fully-connected layer ($\to\num{1024}$, reshaped to $(4,4,4,16))$ to transform the latent $z$ before using 3D deconvolutions (channels: $16\to8\to6$, kernelsize: $2^3$, stride $2^3$) to upsample to \num{4096} particles. While generating a larger PC that matches the input size of $\num{3e4}$ particles would be beneficial to obtain a smoother loss surface, training a network that generates only \num{4096} particles as output is also feasible and will reduce the VAE network size. In the present case, using $\num{3e4}$ particles as an input during training, the network can learn to generate important particles to represent the target PC. Randomly sampling fewer particles from the input would not yield a sufficient representation.%\todo[color=green]{Please reformulate -- better? -- Still correct?}
    \item The INN for the inverse mapping (violet) is built from four Glow coupling blocks~\cite{KingmaDhariw2018_GlowGenerFlowInver1x1Convo} using MLPs with $\to272\to256\to544$ hidden layers as subnets. The hidden layer sizes were chosen to form a bottleneck and reduce to a computationally efficient power of two.
\end{itemize}

The loss function to be minimized during training of this architecture consists of five terms, two connected to the VAE and three for the INN, detailed below:
\begin{itemize}
 \item $L_\mathrm{CD}$: As the reconstruction loss of the VAE, we use the Chamfer's distance (CD)~\cite{FanSuGuibas2016_PointSetGeneratiNetwork3dObject}, it being the most popular differentiable function for comparing PCs the in literature, owing to it being cheap to compute.
 \item $L_\mathrm{KL}$: Regularization of the VAE's latent space is controlled by Kullback-Leibler divergence (KL)~\cite{KingmaWellin2013_AutoEncodingVariatioBayes}.
 \item $L_\mathrm{MSE}$: The resulting spectrum $I'$ of the INNs forward function is compared to the ground truth using mean-squared error $\operatorname{MSE}(I', I)$  
 %{\color{blue}SC: is this loss an established theory or something we created to defind the ground truth? If former, it needs a citation --- JK: It is established theory}
 \item $L_{\mathrm{MMD}(\mathcal{N}, \mathcal{N}')}$, $L_{\mathrm{MMD}(z, z')}$: Both the posterior normal distribution produced by the INN's forward pass $\mathcal{N}`$ and the prior distribution produced by the INN's backward pass $z'$ are evaluated using maximum mean discrepancy (MMD) using an inverse multi-quadratic kernel~\cite{ArdizzKruseWirkerRahnerKothe2018_AnalyInverProblInverNeuraNetwo}.
\end{itemize}
The total loss corresponds to the weighted sum $L$, where, 
%\vspace{-0.1mm}
%\begin{equation}
%\begin{align}
%    L =  L_\mathrm{CD} &+ 0.001\, L_\mathrm{KL} \nonumber\\
%    & + 0.1\,\big(3\, L_\mathrm{MSE}\hspace{-40pt}  &&+ 400\,L_{\mathrm{MMD}(z, z`)}& \nonumber\\
%    &&& + 0.3\,L_{\mathrm{MMD}(\mathcal{N}, \mathcal{N}`)}\big)&
%\label{eq:L}
%\end{align}
%
%\end{equation}
%
%\begin{equation}
 %   L = 1\, L_\mathrm{EMD} + 0.001\, L_\mathrm{KL} + 0.001\,(3\, L^2 + 400\,L_{\mathrm{MMD}(z, z`)} + 300\,L_{\mathrm{MMD}(\mathcal{N}, \mathcal{N}`)} \label{eq:L}
%\end{equation}
\begin{align}
	\hspace{-4pt}L =  L_\mathrm{CD} \hspace{1pt}&+& 0.001\, &L_\mathrm{KL} &&+& 0.3\, &L_\mathrm{MSE} \nonumber\\
	&+&40\,&L_{\mathrm{MMD}(z, z')} &&+& 0.03\,&L_{\mathrm{MMD}(\mathcal{N}, \mathcal{N}')}.
	\label{eq:L}
\end{align}

The numerical values for the coefficients were empirically tuned to optimize performance of the AE reconstruction and the INN forward and backward predictions.

While training on the continuous data stream of non-steady configurations, we employ experience replay (EP)~\cite{Chaudhry2019OnTE} to avoid catastrophic forgetting of earlier simulation time steps while training on later ones.
To integrate with the streaming setup, we implemented this approach to continual learning as a separate component, called training buffer, between the streaming receiver and the training loop. 
This component holds two buffers: a now-buffer, and an EP-buffer. The now-buffer holds the  $N_\mathrm{now}=10$ latest samples received. 
Each time new samples are received, they are prepended to the now-buffer and any element which would then exceed the buffer size is removed and added to the EP buffer. 
The EP buffer holds a maximum of $N_\mathrm{EP}=20$ elements. 
When new samples are to be added to the EP buffer when it is full, a randomly chosen item is removed.
For a single iteration of the training loop, a batch of data is requested from the training buffer.
This batch constitutes the \textit{training batch}.
It is generated by taking $n_\mathrm{now}=4$ random samples form the now-buffer and $n_\mathrm{EP} = 4$ random samples from the EP buffer, yielding a training batch size of $n_\mathrm{now} + n_\mathrm{EP}= 8$.
In order to increase training success, we perform $n_\mathrm{rep}$ iterations of the training loop per single time step from the data stream.
For each iteration, a new training batch is generated from the now- and EP buffer.

Separating the EP schedule from the training loop via our training buffer allows us to control how many batches we iterate per sample time-step produced, as long as we have some leeway to stall the running simulation if need be. This is crucial to allow the optimizer some amount of exploration, which can only happen sequentially and hence a smaller number of training iterations cannot be compensated by the large batch sizes of data-parallel training.

For all training runs, we use the Adam optimizer~\cite{KingmaBa2017_AdamMethodStochastOptimiza} with $\beta_1 = 0.8$, $\beta_2=0.9$, $\epsilon=\num{{}e-6}$ and weight decay $\lambda = \num{2e-5}$. For data parallel training we employ PyTorch's Distributed Data Parallel (DDP) module with its N/RCCL backend. Learning rates are scaled following a square-root rule~\cite{Krizhe2014_OneWeirdTrickParallelConvolutNeural}.
%\begin{equation}
 %   \label{eq::lrscaling}
 %   l_\mathrm{scaled} = l_\mathrm{base} %\sqrt{b / b_0}, 
% \end{equation}
%where $b$ is the used batch size and $b_0=2$ is the fixed reference batch size. 
In order to accelerate learning with large batch sizes, we train layers belonging to the VAE at a learning rate higher by a factor $m_\mathrm{VAE}$ than that of the INN.

\textit{Technical Challenges:} 
The main drawback of CD, used as the VAE's reconstruction loss, is its lacking sensitivity to point density. It also fails to distinguish some point configurations~\cite{AchlioDiamanMitliaGuibas2018_LearnRepreGenerModel3dPoint}.
 Both of these problems are overcome by the earth mover's distance (EMD)~\cite{RubnerTomasiGuibas2000_EarthMoverSDistanceMetricImage} loss, but at much higher computational cost\footnote{We observed about a $4\times$ increase in batch run times when using EMD~\cite{FeydySejourVialarAmariPeyre2019_InterpolBetweenOptimalTransporMmdUsing} compared to a simple implementation of CD.}. Still, a rather compute and memory efficient differentiable implementation of EMD is available in the geomloss python package~\cite{FeydySejourVialarAmariPeyre2019_InterpolBetweenOptimalTransporMmdUsing}, which, instead of relying entirely on PyTorch, internally uses the KeOps library~\cite{CharliFeydyGlauneCollinDurif2021_KernelOperatioGpuAutodiffWithoutMemory, FeydyGlauneCharliBronst2020_FastGeometriLearningSymbolicMatrices} for performance. Unfortunately, the KeOps library is implemented directly in CUDA, and thus cannot be used on AMD GPUs in Frontier. Hence, we were barred from employing EMD loss in our runs on Frontier. Perhaps the community needs a HIP version of the KeOps library.

\subsection{Setup on ORNL Frontier system} 

\DeclareSIUnit\bit{b}
\DeclareSIUnit\byte{B}

\iffalse
We ran the setup on Frontier whose 9472 compute nodes each consist of a 64-core AMD 3rd Gen Epyc 64 core 2 GHz CPU, attached to 512GB of DDR4 memory, as well as of 4 AMD MI250X GPUs with 2 Graphics Compute Dies (GCD) per GPU, i.\,e.\ effectively 8 GCDs on each node visible to applications. Each GCD can be treated as a separate GPU with each GCD consisting of 64 GB of high-bandwidth memory (HBM) (i.e.\ 128GB per AMD MI250X). The CPU and each GCD are connected through Infinity Fabric CPU-GPU. Each node is connected to the network using four HPE Slingshot \SI{200}{GB/s} NICs (\SI{25}{GB/s})\cite{frontier_docs}.
\fi

While distributing nodes exclusively to either PIConGPU or to the MLapp is easier to achieve in Slurm (i.\,e.\, \emph{inter-node} setup in  Fig.~\ref{fig:streamingsolutions}(c) that is supported by our flexible I/O solution), we decide to  use the \emph{intra-node} setup from Fig.~\ref{fig:streamingsolutions}(c) where each compute node is shared between PIConGPU and the MLapp. This strategy has the advantage that data exchange mostly does not need to leave the node, but requires a more specific resource allocation. More specifically, each single compute node no longer runs homogeneous parallel sub-jobs, but has to locally assign resources to heterogeneous applications. For our setup, this local assignment gives 4 Graphics Compute Dies (GCDs) to PIConGPU and 4 GCDs to MLapp. In openPMD and ADIOS2, each reader application decides on its own which remote datasets to load. Here, the distribution of loaded data regions (i.\,e.\ local blocks within the global dataset) is configured such that data is shared within node boundaries.
%The overall workflow is schematically shown in Fig. \ref{fig:schematic-pic-torch-setup}.

\iffalse
\begin{figure}[tb]
    \centering
    \includegraphics[width=\linewidth]{torch_setup_cropped.pdf}
    \caption{\textit{Schematic Setup of PIConGPU+MLapp on a Frontier compute node. Each compute node hosts 4 instances of each PIConGPU and the MLapp. Communication within these parallel processes is realized by their own application logic. Data is exchanged via openPMD and the SST engine of ADIOS2 from PIConGPU to the MLapp.}}
    \label{fig:schematic-pic-torch-setup}
   \vspace{-3mm}
\end{figure}
\fi

Due to the plugin-based structure of PIConGPU, both the particle and radiation input required by MLapp are provided by distinct output plugins of PIConGPU, meaning that two parallel data streams are opened between PIConGPU and the MLapp. 
Each plugin prepares the current time step's data and, upon having all data ready, ADIOS2 will gather the metadata, including remote read addresses to parallel rank 0 and provide the time step to the reader. The reader performs remote reads to its liking and closes the time step, indicating to the writer that the data can now be dropped.

A unique and important requirement of the presented workflow is the ability to not only run but also scale on a large number of Frontier's AMD MI250X GPUs. 
%Software libraries for machine learning on AMD GPUs are maturing as we speak.
As our machine learning model is small enough to fit on a single GCD, parallel training of this model is done using data parallelism, where copies of the model are distributed across GCDs with each copy of the model receiving different chunks of data to train on. Once each model computes its gradients, all the instances of the model must do a collective all-reduce communication to average the gradients.
%Software libraries is a solved problem for most machine learning workflows that are running on NVIDIA GPUs with their mature and stable software stack. Machine learning on AMD GPUs is nascent and brings its own set of challenges while the ROCM software platform becomes more mature. \textcolor{blue}{For SC. This sentence}
We expect the scaling of the workflow to critically depend on the optimization of this all-to-all communication in PyTorch DDP, since communication within the PIConGPU simulation is only between next neighbors. 
%
%Running PyTorch workflows on Frontier requires installing a specific version of PyTorch in either an Anaconda environment or python virtual environment with a corresponding ROCm software module.
%Configuring the environment including resolving dependency conflicts between software packages necessary for the custom fork of ADIOS2~\cite{godoy2020adios} and for PyTorch installation with ROCm. 
%To run PyTorch distributed with slurm, we configured the launch script to run PyTorch with distributed training. 
%PyTorch DDP uses N/RCCL backend for communication needed to perform distributed training across GPUs within and between compute nodes.
%
We experienced a hard limit to scaling in our current setup: The all-to-all communication between PyTorch DDP ranks using the N/RCCL backend hits system limitations on the possible number of open sockets beyond 100 nodes. Potential options to circumvent this limit include using a libfabric backend for N/RCCL or using PyTorch DDP's MPI backend. 
While the N/RCCL backend was able to support PyTorch up to the scale used in this study, we can already conclude that further extensive evaluation of the communication backends within PyTorch will become mandatory to provide a stable base for our near future scaling studies on Frontier. This area of research appears to be still under development, to the best of our knowledge~\cite{dash2023optimizing}.

\iffalse
\textit{2? paragraphs = 1.0 column} \textbf{Klaus, Franz, Jeffrey, Vineeth}
\begin{itemize}
    \item Briefly describe the OLCF's Frontier system with its AMD GPUs, node layout, etc. (only facts relevant for later) -- Done
    \item adaptability to changes in the setup: different number of GPUs for PIC/Torch, future data reduction techniques
    \item Give an overview of how the entire system is connected: PIConGPU distributed over many nodes, ML resides on same nodes, streaming connects and favors inter-node data transfer -- Done
    \item \textit{Plot: simple sketch of setup}
    \item Give estimates of the data flow between PIConGPU and ML
    \item Explicitly state how many epochs are run between each PIC iterations etc. 
    \item Explain in detail how local ML models are combined to a global model. 
\end{itemize}
\fi

\section{Results}

\subsection{Weak-Scaling}

\begin{figure}[tb]
	\centering
	\includegraphics[trim={0 0 0 0.5cm},clip,width=.4\textwidth]{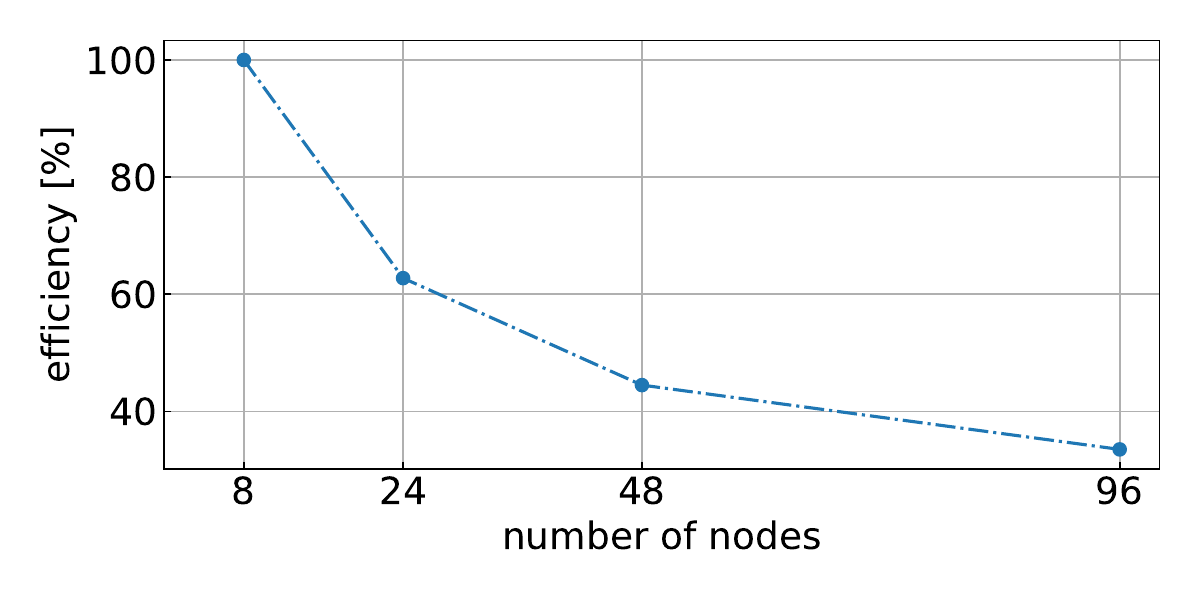}
	\vspace{-2mm}
	\caption{\small{Frontier weak scaling of the in-transit training from 32 to 384 GCDs (8 to 96 nodes). Measurements are single-batch times averaged over multiple runs and over iterations within runs after removal of $>4\sigma$ outliers.}}
	\label{fig:weakscaling}
	\vspace{-5mm}
\end{figure}

When training any kind of statistical model, in our case a deep neural network, one has to ensure that for each characteristic of the domain one hopes to learn, there are sufficient examples available in the dataset. When the dataset is provided in-transit, redundancy must still be built into the setup, either by using a large simulation volume or running multiple independent simulations concurrently, to allow the network to learn repeated patterns of the simulations and thereby distinguish relevant domain features, such as physical relationships, from noise. For this reason, in this work, we focus on a weak-scaling study of the proposed pipeline, as we anticipate this to be the most likely use-case.
To evaluate parallel scaling as well as learning scaling, we performed simulations with in-transit training from 8 up to 96 nodes.
%, with half of the available GCDs of each node assigned to the simulation and the other half to the training, this corresponds to 32 to 384 GCDs for each. 
The lower limit of 8 nodes is set by physical size constraints of the simulation. Fig.~\ref{fig:weakscaling} shows the weak scaling behavior of the pipeline on Frontier. We define efficiency as the ratio of runtime at a specific size over runtime of the smallest size. Before averaging the times per training batch, we removed outliers more than four standard-deviations from the mean, as we observed single batches taking more than $100\times$ the mean time on Frontier.

While ideal scaling should correspond to 100\,\% efficiency across all sizes, communication overheads reduce efficiency in practice. Here, the efficiency of in-transit training reaches around 35\,\% at 96 nodes.
We attribute the low parallel efficiency to two main reasons. 
%One is the inevitable all-to-all communication between PyTorch ranks taking place to average gradients during each backward pass and to obtain a synchronized result on all ranks, accounting for a deficit of $\sim30\,\%$, which is still a respectable result achieved through compute--communication overlap in PyTorch DDP. 
The first reason is the inevitable all-to-all communication between PyTorch ranks that takes place to average the gradients during each backward pass and obtain a synchronized result on all ranks, which accounts for a deficit of $\sim30\,\%$, which is still a respectable result achieved by overlapping computations and communication in PyTorch DDP.
The second reason is the computation of the two MMD loss terms, which amount to matrix dot products with data distributed across all ranks. 
As torch (less than version 2.2) did not offer a distributed primitive and the matrices involved are relatively small, even at the largest achieved scale, we opted for a naive implementation which replicates work between nodes. The global communication required here\footnote{\texttt{torch.distributed.all\_gather\_into\_tensor}} breaks the torch computational graph, i.e.\ synchronizes graph execution with host code at the invocation site. To our knowledge there is no high-level API available to stitch these operations into the graph to allow them to be executed asynchronously to both the host and each other, i.e no enqueueing of host functions or events.
%
%\todo[color=green]{What do we want to write about the learning rate and buffer size scaling?}

\iffalse
\begin{itemize}
    \item Motivate that we focused on weak scaling only so that the ML model always learns on the same amount of data and that results will not be influenced by a lack of data (and other reasons). 
    \item State that we preformed simulations from 8 nodes to \textbf{96} nodes.
    \item Plot: Show weak scaling plot
    \item Discuss how increasing the number of resources effected the duration of the simulations. 
    \item Provide technical details (if possible) on the reduction of performance observed. 
    \item State in detail the distribution of compute power to each of the 3 parts (PIC, streaming, ML) 
    \item Put weak-scaling measurement into context
    \item Discuss what the main limiting factor(s) are
    \item Conclude how this limits this method of an artificial scientist
    \item Provide possible solutions how to overcome these bottlenecks     
\end{itemize}
\fi

\subsubsection{Scaling Model Training} The performed scaling runs, training with a batch size of $n_\mathrm{now}+n_\mathrm{rep}=8$ per GCD, correspond to total batch sizes of 256 to 3072 on 32 to 384 GCDs, respectively. 

%(no problem here, just did not do pretraining in the end) Our preferred method to stabilize training at scale is using a pre-trained model (\textit{see Section~\ref{machine_learning}}). While pursuing this approach, we ran into challenges as the pre-trained model that worked as expected on large scale simulations on a smaller local system (NVIDIA GPU-based), did not produce correct results on Frontier; this requires further evaluation.
%For similar reasons, the upcoming discussions on the model's potential, shown in Fig.~\ref{figPhysicalResult} is based on results from NVIDIA GPUs (\textit{see Section~\ref{sec:phyMLdiscussion}}).

%{\color{red}@SC: this first point is just very blunt, as we discussed, not sure how it should be phrased or if it should be in at all.} Our preferred method to stabilize training a scale started with a pre-trained model, for reason detailed before {\color{blue}SC: in which section? Cross-cite it here}. In this pursuit, we encountered difficulties getting a model trained on a smaller local system, performing correctly on Frontier. We were able to verify that our pre-trained model performed as expected on larger scale simulations by evaluating the model offline on generated on Frontier on a local system.

In our scaling runs, we explored training from scratch at large scale, sampling up to $n_\mathrm{rep}=96$ batches per sample consumed from the stream and found learning success up to about $n_\mathrm{rep}=48$. We settled on a base-learning rate of $l_\mathrm{base}=$\num{{}e-6}, 
scaled according to the square-root scaling rule~\cite{Krizhe2014_OneWeirdTrickParallelConvolutNeural} depending on the actual number of GCDs.
We observe a training speed up with larger batch sizes. That is, the increased learning rate overcompensates the training slow down due to larger batch sizes.
Furthermore, it became evident that the three losses connected to the INN converged best at  lower learning rates, while the VAE only found better minima at the highest learning rate, showing that separate learning rates $l_\mathrm{VAE}$ and $l_\mathrm{INN}$ for the encoder/decoder block and the inversion block respectively need to be applied at large scales.

Within the scope of this study, we have to conclude that for in-transit training at very large batch sizes, hyper parameter studies do not transfer from small-scale experiments and thus have to be performed at scale. To enable in-transit learning at the scales we tested or beyond, comprehensive studies of the relations between the block learning rates $l_\mathrm{VAE}$ and $l_\mathrm{INN}$, batch sizes, and maybe even loss weights in Eq.~\eqref{eq:L} have to be performed. 
Exploring this parameter space for the large scale training requires the full
study to be systematically performed at these scales (up to 384 GCDs), either by
employing a streamed simulation or the same training data stored on disk. We
argue that the latter is challenging at scale due to the lack of disk capacity
and bandwidth. Efficient approaches to autotuning and optimization will be
indispensable for this next step.

\subsection{Quantifying the predictive capabilities of the machine learning method} \label{sec:phyMLdiscussion}

\begin{figure}[tb] 
\centering
\includegraphics[width=0.4\textwidth]{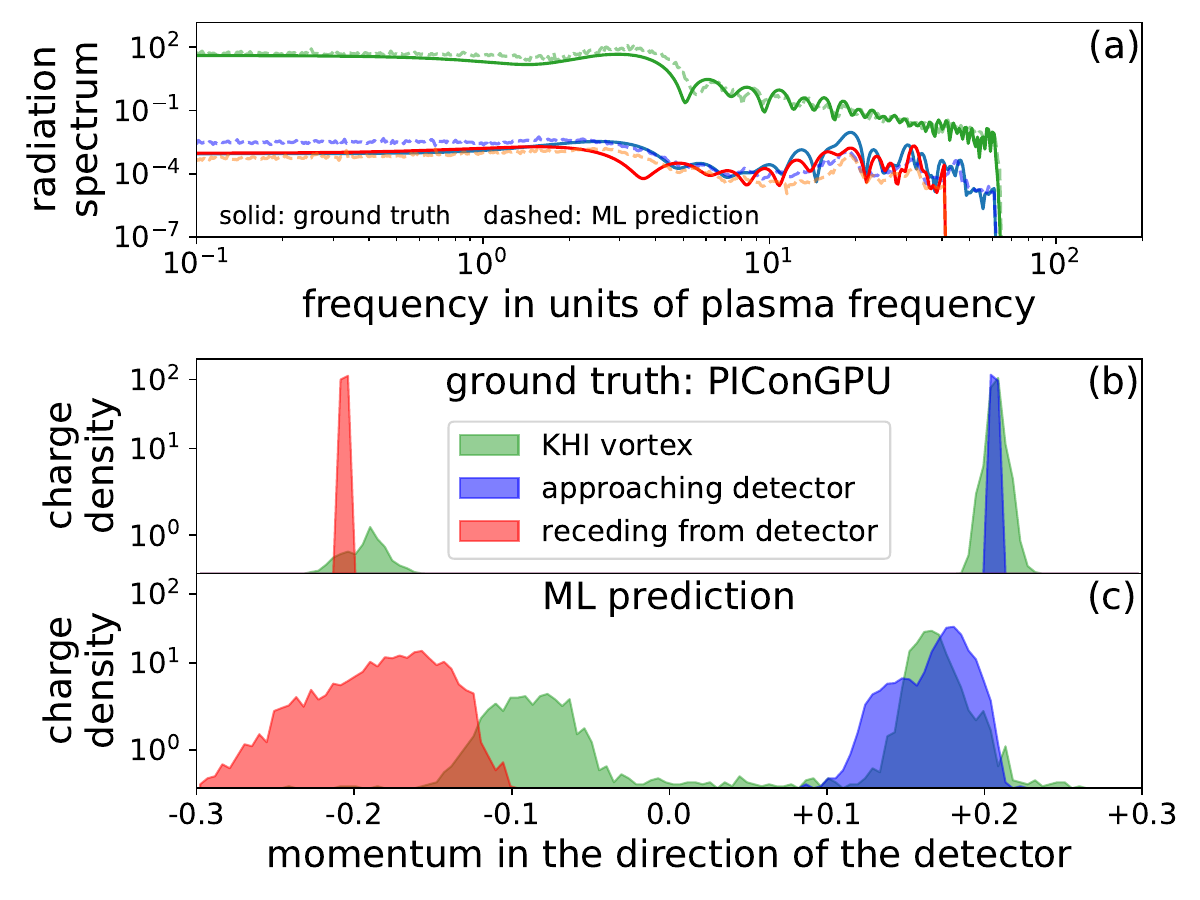}
\caption{\small{Comparing the ground truth data from PIConGPU with the prediction of the ML model on a selected example sub-volume. Blue \& red are undisturbed plasma streams approaching or receding from the detector, green depicts the KHI vortex regions. (a) radiation spectra, (b) ground truth and (c) predicted momentum distribution. The agreement is good enough to unambiguously classify the region of origin of the plasma instability based on the momentum distribution predicted by the radiation spectra.}} 
\label{figPhysicalResult}
\vspace{-0.2in}
\end{figure}

To evaluate the performance of the trained model, we invert the radiation spectra back to the original momentum distribution, focusing on the momentum component $p_x$ in the following discussion (see Fig.~\ref{figPhysicalResult}).
As this inversion is an unsolved problem in physics, we do not expect good agreement, but would be satisfied with partial reconstruction of the distribution.

Given spectra from the bulk of the plasma, the model reconstructs the momentum distribution quite well (blue and red Fig.~\ref{figPhysicalResult}(c)). 
The mean of the predicted momentum agrees with the distribution from PIConGPU (Fig.~\ref{figPhysicalResult}(b)). 
We also observe that the network learned a fundamental aspect of special relativity: the Doppler shift, to distinguish between plasma streams approaching and receding from the detector, since it reproduces the correct cutoff frequency when predicting spectra from particle data (Fig.~\ref{figPhysicalResult}(a) dotted lines).

The momentum prediction from the KHI shear surface was expected to be more difficult  
since particles from both directions cross in the shear surface, resulting in two separate and distinct particle populations.
The ML reconstruction consistently predicts these two distinct populations.
This is consistent with the interpretation that the INN has learned unsupervised that the vortex region undergoes more acceleration and thus emits more intense radiation than the bulk of the plasma.
However, the momentum prediction is less accurate: 
the mean of (one of) these populations often does not match the PIConGPU data (Fig.~\ref{figPhysicalResult}(b vs. c) green).
We suspect that this is partly due to our choice of region boundaries on the shear surface, which results in a much lower sampling rate of one flow direction compared to the other, leading to a lower accuracy in predicting the mean of the lower peak of the momentum distribution. 
It is also possible that this problem arises from the VAE's decoder having difficulties reconstructing discontinuous distributions. 
%Nevertheless, the successful prediction of shear regions suggests that the encoder extracts from the input data crucial features to discern whether a region contains a vortex and encodes them in latent space. 
Nevertheless, the successful prediction of shear regions indicates that the
encoder extracts crucial features from the input data to recognize whether a
region contains a vortex and encodes them in latent space.

The model clearly learned to partition the latent space into regions for different flow
directions and vortex regions, which both the encoder and the
inversion network learned to map to. 
%These features, extracted using loss
%functions that did not impose physical rules, do in principle allow a simple, almost linear,
%classifier to predict the physical regimes, which presents possibility to verify the
%extraction of physical rules
These features are extracted using loss functions that do not impose physical rules and still encode physical properties of the system.
Hence, in principle, they allow a simple, almost linear classifier to predict physical regimes. Evaluating such a classifier quantifies successful extraction of underlying physical rules.

We observed that main issue of our model is suboptimal reconstructions of vortex momenta,
which we attribute to shortcomings of our decoder architecture,
despite which the ML prediction still allows to visually identify KHI regions.
Previous work on PC generation~\cite{pmlr-v80-achlioptas18a}  observed that the latent space of an AE may contain more information than the decoder can reconstruct with satisfactory fidelity and that more sophisticated generative models, such as GANs or flows, can learn to produce more detailed reconstructions when conditioned on the same latent space.
%While we do not yet have a perfect reconstruction of the KHI particle parameters, the identification of regions of instability is a major step forward. 

While the prediction accuracy for the inversion from radiation to momentum is still limited, our workflow shows promising potential for automated learning of physical relationships from in-transit data of large-scale simulations and the identification of regions of instability is already a major step forward. 
This identification by the network may be obvious to a domain expert, but the \textit{Artificial Scientist} learned the distinction unsupervised. 
The work has demonstrated, as a proof of principle, that it \textit{is} possible to learn correlations from a physics simulation on-the-fly.
\textbf{This is a promising first result towards large-scale in-transit learning for a non-steady state processes.}

%\sunita{we need to squish this, else we are going to have at least one reviewer that says - do this and then re-submit :D. there are 4 bullets, let us give just 2 bullets and keep it generic so we also don't give away our ideas}
Future work will explore more sophisticated network components such as \textbf{(a)} Deeper INNs with more powerful subnet-architecutres~\cite{DinhSohl-DBengio2016_DensityEstimatiUsingRealNvp}, \textbf{(b)} a wealth of more sophisticated PC decoder networks to generate higher-fidelity depictions of predicted particle distributions~\cite{pmlr-v80-achlioptas18a}, and \textbf{(c)} more complex encoders architectures~\cite{QiSuMoGuibas2017_PointnetDeepLearningPointSets3d}, ideally bringing contrastive learning approaches~\cite{CaronTouvroMisraJegouJoulin2021_EmergingPropertiSelfSupervisVisionTransfor} to point clouds to learn better latent representations.
\iffalse
\begin{itemize}
 \item Deeper INNs with convolutional subnets~\cite{DinhSohl-DBengio2016_DensityEstimatiUsingRealNvp} to allow for a more non-linear transformation while keeping the number of trainable parameters low. 
 \item More sophisticated decoder networks to generate higher-fidelity depictions of predicted particle distributions~\cite{pmlr-v80-achlioptas18a}.
 \item Encoders incorporating point-transformer blocks~\cite{QiSuMoGuibas2017_PointnetDeepLearningPointSets3d} and a deeper network around the bottleneck to better extract latent information from point-features.
 \item Adapting contrastive learning approaches~\cite{CaronTouvroMisraJegouJoulin2021_EmergingPropertiSelfSupervisVisionTransfor} to point clouds in order to learn better latent representations.
\end{itemize}
\fi
\section{Conclusion}
By training a machine learning model on-the-fly on a PIConGPU simulation of the Kelvin-Helmholtz instability, where data is streamed via openPMD with ADIOS2, we demonstrate a realization of the \textit{Artificial Scientist} solving an ill-posed inverse problem from plasma physics.
This allows us to circumvent the lack of adequate disk capacity and bandwidth, often a bottleneck in large-scale systems, by enabling data to reside in-memory and be distributed via network between the nodes. Scientifically, as a proof of principle, this work also demonstrates that it is possible to learn correlations from a physics simulation on-the-fly.
The presented workflow is scalable from local clusters to Top-1
supercomputers and is tested across GPU vendors. This is possible due to PIConGPU's foundation on the abstraction layer alpaka and due to the availability of various ports of PyTorch. 
On full Frontier, we demonstrate streaming at scale using openPMD as well as performance studies at scale for general plasma physics.

\iffalse
\begin{itemize}
    \item Summarize that we presented the first demonstrate of an in-transit simulation analysis aka \textit{Artificial Scientist}
    \item Summarize that the principle workflow has been demonstrated to scale (well?) and that simple predictions have been reached within a very short training time. 
    \item Provide an outlook that for future campaigns of this kind, a number of improvements (\textbf{provide list}) is needed.
    \item Conclude that these findings are promising for overcoming the soon-to-be limit of IO storage in large-scale exploratory simulations. 
\end{itemize}
\fi

%There are more complex radiation inversion problems, where the total radiation signal does not decompose into radiation from spatially localized regions. For these kind of problems a different approach needs to be chosen.
%Therefore, a solution that solely relies on independent algorithms each operating on a spatially localized region without communication will by the nature of the problem, not work.

\section*{Acknowledgments}
\footnotesize{
The authors thank Patrick Stiller and Nico Hoffmann for fruitful discussions.
This research used resources of the Oak Ridge Leadership Computing Facility at the Oak Ridge National Laboratory, supported by the Office of Science of the U.S. Department of Energy under Contract No. DE-AC05-00OR22725.
This research was supported by the Helmholtz Association Initiative and Networking Fund in the frame of Helmholtz AI, 
by Laserlab-Europe EU-H2020 GA no. 871124 and the European Union through Grant Agreement No. 101093261 (Plasma-PEPSC). 
The support work from the ADIOS team was supported by the U.S. Department of Energy, Office of Science, Office of Advanced Scientific Computing Research, Scientific Discovery through Advanced Computing (SciDAC) program, under the “RAPIDS Institute”.
}
\bibliography{bibliography}
\bibliographystyle{ieeetr}

%\begin{thebibliography}{00}
%\bibitem{b1} G. Eason, B. Noble, and I. N. Sneddon, ``On certain integrals of Lipschitz-Hankel type involving products of Bessel functions,'' Phil. Trans. Roy. Soc. London, vol. A247, pp. 529--551, April 1955.
%\end{thebibliography}
%\vspace{12pt}
%\color{red}
%IEEE conference templates contain guidance text for composing and formatting conference papers. Please ensure that all template text is removed from your conference paper prior to submission to the conference. Failure to remove the template text from your paper may result in your paper not being published.

\end{document}

%% file: tikzFigsPicks.tex
\input{hsl.tex}

\definecolorHsl{producer}{236}{.3}{0.8}
\definecolorHsl{consumer}{0}{.3}{.7}
\definecolorHsl{nodeBG}{0}{0}{.9}
\definecolorHsl{stream}{40}{.9}{.4}

\definecolorHsl{encoder}{80}{.5}{.6}
\definecolorHsl{decoder}{180}{.5}{.5}
\definecolorHsl{inverter}{300}{.5}{.4}
\definecolorHsl{latent}{80}{.5}{.3}

\definecolorHsl{microstate}{200}{.5}{.6}
\definecolorHsl{microstateApprox}{200}{.5}{.5}
\definecolorHsl{physicalRed}{120}{.5}{.4}

\newlength\lenDashLong
\newlength\lenDashShort
\newlength\lenDot
\newlength\lenGap
\newlength\prodConsWidth
\newlength\distance
\newlength\rowskip

\tikzset{
 actor/.style={draw, rectangle,align=center, rounded corners=1.5ex},
 actorDomain/.style={fill=gray, draw, rectangle,align=center,},
 process/.style={draw, rectangle,align=center},
 learnTag/.style={font=\color{red}\bf,text opacity=1, fill opacity=.5,fill=white,
 node distance=.5ex,anchor=south west},
 flow/.style={thick, -{Latex[length=1.5ex,width=1ex]}},
 comm/.style={very thick,green!80!black,{Latex[length=.2ex,width=.5ex]}-{Latex[length=.2ex,width=.5ex]}},
 rect/.style={inner sep=0,minimum width=#1,minimum height=#1,draw=black},
 prodBox/.style={fill=producer,minimum height=2em,minimum width=\prodConsWidth},
 prod/.style={fill=producer,rect={.5cm},draw=black},
 consBox/.style={fill=consumer,minimum height=2em,minimum width=\prodConsWidth},
 cons/.style={fill=consumer,rect={.5cm},draw=black},
 node/.style={fill=nodeBG,draw=black},
 microstate/.style={draw=microstate,modelDraw,solid},
 microstateApprox/.style={draw=microstateApprox,modelDraw,dash pattern=on
 \lenDot
 off \lenGap,solid}, % solid overrides dash pattern (after)
 physicalRed/.style={draw=physicalRed,modelDraw,dash pattern=on \lenDashLong off
 \lenGap,solid},
 modelDraw/.style={line width=.5ex,},
 encoder/.style={draw=encoder,modelDraw,
 dash pattern=on \lenDashLong off \lenGap on \lenDot off \lenGap,solid},
 decoder/.style={draw=decoder,modelDraw,
  dash pattern=on \lenDashLong off \lenGap on \lenDot off \lenGap on \lenDot off \lenGap on \lenDot off \lenGap,solid},
 inverter/.style={draw=inverter,modelDraw,dash pattern=on \lenDashLong off \lenGap on \lenDashShort
 off \lenGap,solid},
 latent/.style={draw=latent,modelDraw,
  dash pattern=on \lenDashLong off \lenGap on \lenDot off \lenGap on \lenDot off \lenGap,solid},
}
\tikzset{
 pics/hddMock/.style 2 args={code={
  \begin{scope}[anchor=center]
   \coordinate(left) at (-.5#1,0);
   \coordinate(right) at (.5#1,0);
   \coordinate(low1) at (-.5#1,-#2);
   \coordinate(low2) at (-.5#1,-2*#2);
   \coordinate(low3) at (-.5#1,-3*#2);
   \coordinate(low4) at (-.5#1,-4*#2);
   \draw (left) to [bend left=90] (right) to [bend left=90] (left)
   -- (low1)
   to [bend right=90] (low1 -| right) -- (right);
   \draw (low1) -- (low2) to [bend right=90] (low2 -| right) -- (right);
   \draw (low2) -- (low3) to [bend right=90] (low3 -| right) -- (right);
   \draw (low3) -- (low4) to [bend right=90] (low4 -| right) -- (right);
   \node[font=\small, anchor=center] at (0,0) {storage};
  \end{scope}
 }},
 nnMock node/.style={draw,circle,minimum width=.7*#1,inner sep=0},
 pics/nnMock/.style 2 args={code={
   \foreach \x in {-1,...,1}{
    { \node[nnMock node={#2}] (a \x) at (-#1, \x*#2) {}; };
   };
   \foreach \x in {-2,...,2}{
    { \node[nnMock node={#2}] (b \x) at (0, \x*#2) {}; };
    \foreach \i in {-1,...,1}{
     \draw (a \i) -- (b \x);
    };
   };
   \foreach \x in {0,1}{
    { \node[nnMock node={#2}] (c \x) at (#1, \x*#2-.5*#2) {}; };
    \foreach \i in {-2,...,2}{
     \draw (b \i) -- (c \x);
    };
   };
 }},
 pics/domainDecompostionComm/.style n args={3}{code={
   \foreach \x in {0,...,#2}{
    \foreach \y in {0,...,#2}{
     \node[actorDomain,rectangle,inner sep=0,minimum width=#1*.5,minimum
     height=#1*.5,anchor=north west,#3, draw=black,sharp corners] (n \x \y) at (#1*\x, #1*\y)
      {};
     \pgfmathparse{\x > 0}
     \if\pgfmathresult1
      \pgfmathtruncatemacro{\xprev}{\x -1}
      \draw[comm] (n \x \y) -- (n \xprev \y);
     \fi
     \pgfmathparse{\y > 0}
     \if\pgfmathresult1
      \pgfmathtruncatemacro{\yprev}{\y -1}
      \draw[comm] (n \x \y) -- (n \x \yprev);
     \fi

    };
   };
 }},
 pics/domainDecompostion/.style n args={3}{code={
   \foreach \x in {0,...,#2}{
    \foreach \y in {0,...,#2}{
     \node[actorDomain,rectangle,inner sep=0,minimum width=#1*.5,minimum
     height=#1*.5,anchor=north west,#3,draw=black,sharp corners] (n \x \y) at (#1*\x, #1*\y)
      {};
    };
   };
 }},
}

\newcommand\internodestream{
  node[midway, anchor=center,font=\color{gray}, scale=\nodescale] (arrow inter) {
   \begin{tikzpicture}
     \def\xn{0}
     \foreach \yn in {0,...,1}{
      \path[node] (\xn*\nxscale,\yn*\nyscale) rectangle +(\nxsize,\nysize);
      \begin{scope}[local bounding box={n \xn_\yn}]
       \foreach \x in {0,...,1}{
        \foreach \y in {0,...,1}{
         \node[prod,anchor=center,] (r \x_\y) at
         (\xn*\nxscale+\x*\rxscale+\rxscale/2,
         \yn*\nyscale+\y+.5)
         {};
        };
       };
      \end{scope}
      \def\xn{1}
      \path[node] (\xn*\nxscale,\yn*\nyscale) rectangle +(\nxsize,\nysize);
       \begin{scope}[local bounding box=n \xn_\yn]
       \foreach \x in {0,...,1}{
        \foreach \y in {0,...,1}{
         \node[cons,anchor=center,] (r \x_\y) at
         (\xn*\nxscale+\x*\rxscale+\rxscale/2,
         \yn*\nyscale+\y+.5)
         {};
        };
       };
      \end{scope}
      \draw[inter stream,] (n 0_\yn.east) -- (n 1_\yn);
     };
   \end{tikzpicture}
   }
}

\newcommand\mixedstream{
  node[pos=.825, anchor=center,font=\color{gray},scale=\nodescale] (arrow mix) {
   \begin{tikzpicture}
    [
     pics/mixed node/.style={code={
      \path[node] (0,0) rectangle +(\nxsize,\nysize);
      \begin{scope}[local bounding box={n ##1}]
       \def\x{0}\def\y{0}
       \node[prod,anchor=center,] (r \x_\y) at
        (\x*\rxscale+\rxscale/2,\y+.5) {};
       \def\x{1}\def\y{0}
       \node[prod,anchor=center,] (r \x_\y) at
        (\x*\rxscale+\rxscale/2,\y+.5) {};
       \def\x{0}\def\y{1}
       \node[prod,anchor=center,] (r \x_\y) at
        (\x*\rxscale+\rxscale/2,\y+.5) {};
       \def\x{1}\def\y{1}
       \node[cons,anchor=center,] (r \x_\y) at
        (\x*\rxscale+\rxscale/2,\y+.5) {};
       \draw[stream,
         line width=.4cm, -{Latex[length=\nskip cm,width=.8cm]}]
        (r 0_1.east) -- (r 1_1);
      \end{scope}
     }},
     pics/cons node/.style={code={
      \path[node] (0,0) rectangle +(\nxsize,\nysize);
      \begin{scope}[local bounding box={n ##1}]
       \foreach \x in {0,...,1}{
        \foreach \y in {0,...,1}{
         \node[cons,anchor=center,] (r \x_\y) at
          (\x*\rxscale+\rxscale/2,\y+.5) {};
         };
        };
      \end{scope}
     }},
    ]
    \pic at (0,0) {mixed node={0 0}};
    \pic at (\nxscale,0) {mixed node={1 0}};
    \pic at (0,\nyscale) {mixed node={0 1}};
    \pic at (\nxscale,\nyscale) {cons node={1 1}};
    \draw[inter stream,] (n 0 1.east) -- (n 1 1);
   \end{tikzpicture}
   }
}

\newcommand\intranodestream{
  node[pos=.175, anchor=center,font=\color{gray},scale=\nodescale] (arrow intra) {
   \begin{tikzpicture}
    \foreach \xn in {0,...,1}{
     \foreach \yn in {0,...,1}{
      \path[node] (\xn*\nxscale,\yn*\nyscale) rectangle +(\nxsize,\nysize);
      \def\x{0}
      \begin{scope}[local bounding box={n \x_\yn}]
        \foreach \y in {0,...,1}{
         \node[prod,anchor=center,] (r \x_\yn) at (\xn*\nxscale+\x*\rxscale+\rxscale/2,
         \yn*\nyscale+\y+.5)
         {};
        };
      \end{scope}
      \def\x{1}
       \begin{scope}[local bounding box=n \x_\yn]
        \foreach \y in {0,...,1}{
         \node[cons,anchor=center,] (r \x_\yn) at (\xn*\nxscale+\x*\rxscale+\rxscale/2,
         \yn*\nyscale+\y+.5)
         {};
        };
      \end{scope}
      \draw[inter stream,] (n 0_\yn.east) -- (n 1_\yn);
      };
     };
   \end{tikzpicture}
   }
}

\newcommand\figStreamSolutions
{
 \setlength\rowskip{-7em}
 \newlength\subFigCaptionDist
 \setlength\subFigCaptionDist{.1em}
 \begin{tikzpicture}
  [
   bigStreamP/.style={line width=##1, -{Latex[length=1.5em,width=##1/.6]},
   stream},
   bigStream/.style={bigStreamP=1em},
   actor/.append style={node distance=3em},
   caption/.style={anchor=north west,
    execute at begin node={\begin{minipage}{\linewidth-3em}},execute at end
    node={\end{minipage}}, font=\small}
  ]

  \pgfdeclarelayer{back}
  \pgfdeclarelayer{front}
  \pgfsetlayers{back,main,front}

  \node[actor,anchor=west, prodBox] (prod) at (0,0) {producer};
  \node[actor, anchor=east,consBox] (cons) at (\linewidth,0) {consumer};
  \draw[bigStream] (prod.east) -- (cons.west)
   node[midway,font=\small] (stream) {\color{white}streaming}
   node[below,midway,font=\small] {$\mathcal{O}(\SI{}{PB/s})$};

  \node[above=2.2em of prod, anchor=center] {\includegraphics[width=5em]{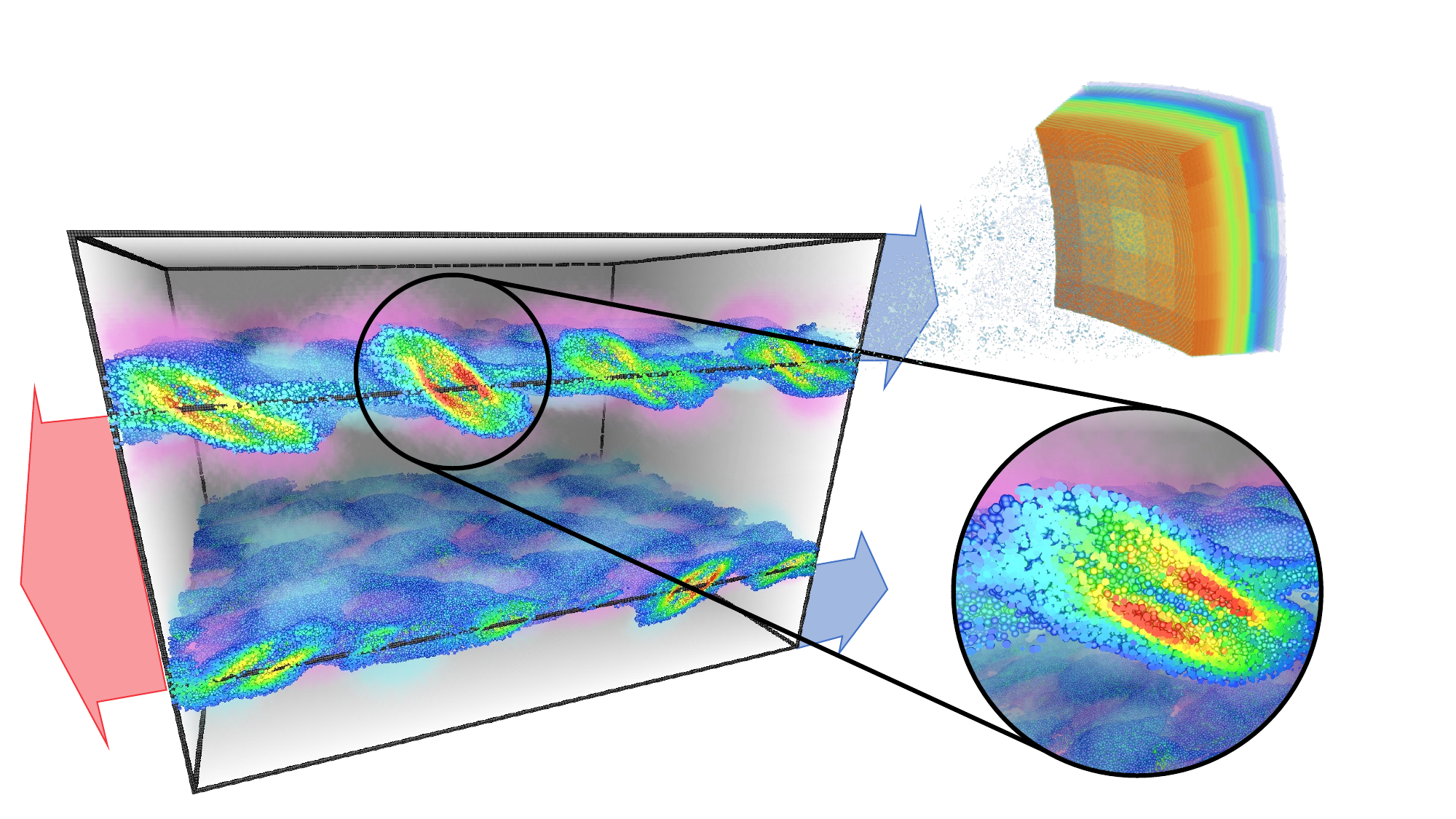}};

  \pic[above=2.2em of cons,local bounding box=nn,scale=.6,transform shape] {nnMock={.7}{2.5ex}};

  \pic[above= 3em of stream,local bounding box=hdd] {hddMock={1}{.2}};
  \draw[dashed, flow] (prod.east) .. controls +(.9,+.2) and ($(hdd.west)
  -(.9,+.2)$) .. (hdd.west) node[midway,above,sloped,font=\small] {$\sim\SI{}{10TB/s}$};
  \draw[dashed, flow] (hdd.east) .. controls +(.9,-.2) and ($(cons.west)
  -(.9,-.2)$) .. (cons.west);

  \draw[red,very thick,] (hdd.north east) -- (hdd.south west);
  \draw[red,very thick,] (hdd.south east) -- (hdd.north west);

  \node[fit=(prod)(cons)] (img) {};
  \node[caption] at ($(img.south west) +(2em,-\subFigCaptionDist)$)
  {(a) stream via network/in-memory, avoid storage};

  \node[actor,anchor=west, prodBox, below=of prod] (prod) {producer};
  \node[actor, anchor=east,consBox,below=of cons] (cons) {consumer};
  \path (prod) -- (cons)
  node[process, midway, fill=encoder] (reduction) {reduction};

  \draw[bigStream] (prod.east) -- (reduction)
   node[below,pos=-.1,anchor=north west,font=\small] {$\mathcal{O}(\SI{}{PB/s})$};
  \draw[bigStreamP=.4em] (reduction) -- (cons.west)
   node[below,pos=-.1,anchor=north west,font=\small] {$\sim\SI{}{30TB/s}$};

  \node[fit=(prod)(cons)] (img) {};
  \node[caption] at ($(img.south west) +(2em,-\subFigCaptionDist)$)
  {(b) reduction of produced data before or in stream};

  \setlength\rowskip{-6em}

  \node[actor,anchor=west,align=center,font=\small, below=of prod] (prod)
  {Distributed\\Producer\\[1em]
   \begin{tikzpicture}
    \pic[local bounding box=cons dd] {domainDecompostion={2ex}{3}{fill=producer}};
   \end{tikzpicture}
  };

  \node[actor,anchor=east,align=center,font=\small, below=of cons] (cons)
  {Distributed\\Consumer\\[1em]
   \begin{tikzpicture}
    \pic[local bounding box=cons dd] {domainDecompostion={2ex}{3}{fill=consumer}};
   \end{tikzpicture}
  };

  \draw[line width=2em, -{Latex[length=2em,width=2em/.6]}, stream] (prod.east) --
  (cons.west) 
  coordinate[pos=.1] (stream west)
  coordinate[pos=.9] (stream east)
  node[pos=.01,anchor=west,] {\color{white}work distrib. balances BW};

  \def\rxscale{1.2}
  \def\nskip{0.5}
  \pgfmathsetmacro\nxsize{\rxscale*2}
  \pgfmathsetmacro\nxscale{\nxsize+\nskip}
  \def\nysize{2}
  \pgfmathsetmacro\nyscale{\nysize+\nskip}
  \tikzset{
   inter stream/.style={stream,
    line width=\nysize*.6 cm, -{Latex[length=\nskip cm,width=\nysize cm]}}
  }

  \def\nodescale{.4}
  \path ($(prod.south west) -(0,3.6em)$) -- ($(cons.south east) -(0,3.6em)$)
  \intranodestream
  \internodestream
  \mixedstream
  ;

  \node[below=1ex of arrow inter] {inter node};
  \node[below=1ex of arrow mix] {mixed};
  \node[below=1ex of arrow intra] (intraNodeLbl) {intra node};

  \node[fit=(prod)(cons)(intraNodeLbl)] (img) {};
  \node[caption] at ($(img.south west) +(2em,-\subFigCaptionDist)$)
  {(c) producer and consumer are distributed across nodes (gray boxes), arranged
  to balance bandwidth requirements with availablity};

  \begin{pgfonlayer}{back}
   \fill[shading = axis,top color=stream,middle color=stream!10!white]
    (stream west) to[out=-20, in=20] (arrow intra.south west) -- (arrow
    mix.south east) to[out=160, in=-160] (stream east) ;
  \end{pgfonlayer}

 \end{tikzpicture}
}

\newcommand\figModelWorkflow
{
 \begin{tikzpicture}
  [
   every node/.append style={minimum width=5em,minimum height=1.5em,node
   distance=1em},
   frame/.style={draw, thick,inner sep=1ex},
   flow/.append style={stream,line width=1ex, -{Latex[length=1.5ex,width=2.5ex]}}
  ]
  \node[actor,physicalRed,](rad 2) {radiation $I$};
  \node[draw, inverter,below=of rad 2](inn 2) {inversion};
  \node[actor,latent,below=of inn 2](z 2) {latent $z$};
  \node[draw, decoder,below=of z 2](decoder 2) {decoder};
  \node[actor,microstateApprox,below=of decoder 2](particlesOut
  2) {particles $\mathcal{D}'$};
  \draw[flow] (rad 2) -- (inn 2);
  \draw[flow] (inn 2) -- (z 2);
  \draw[flow] (z 2) -- (decoder 2);
  \draw[flow] (decoder 2) -- (particlesOut 2);

  \node[actor,microstate,right=3em of rad 2](particlesIn 3) {particles $\mathcal{D}$};
  \node[draw, encoder,below=of particlesIn 3](encoder 3) {encoder};
  \node[actor,latent,below=of encoder 3](z 3) {latent $z$};
  \node[draw, decoder,below=of z 3](decoder 3) {decoder};
  \node[actor,microstateApprox,below=of decoder 3](particlesOut
  3) {particles $\mathcal{D}'$};
  \draw[flow] (particlesIn 3) -- (encoder 3);
  \draw[flow] (encoder 3) -- (z 3);
  \draw[flow] (z 3) -- (decoder 3);
  \draw[flow] (decoder 3) -- (particlesOut 3);

  \node[actor,microstate,right=3em of particlesIn 3](particlesIn 1) {particles $\mathcal{D}$};
  \node[draw, encoder,below=of particlesIn 1](encoder 1) {encoder};
  \node[actor,latent,below=of encoder 1](z 1) {latent $z$};
  \node[draw, inverter,below=of z 1](inn 1) {surrogate};
  \node[actor,physicalRed,below=of inn 1](rad 1) {radiation $I$};
  \draw[flow] (particlesIn 1) -- (encoder 1);
  \draw[flow] (encoder 1) -- (z 1);
  \draw[flow] (z 1) -- (inn 1);
  \draw[flow] (inn 1) -- (rad 1);

  \node[frame, fit=(particlesIn 1)(rad 1),align=center,] (radiation) {};
  \node[frame, fit=(particlesOut 2)(rad 2),align=center,] (inversion) {};
  \node[frame, fit=(particlesOut 3)(particlesIn 3),align=center,] (compression) {};
  \node[below=1ex of radiation,align=center]  {(c) physical\\forward\\prediction};
  \node[below=1ex of inversion,align=center]  {(a) inversion\\of physical\\reduction};
  \node[below=1ex of compression,align=center]{(b) compression\\--\\decompression};

 \end{tikzpicture}
}

\newcommand\figModelIOOutBW
{
 \setlength\distance{1.5em}
 \newlength\flowwidth
 \setlength\flowwidth{.1ex}
 \resizebox{\linewidth}{!}{%
 \begin{tikzpicture}[
   node distance=\distance, every node/.style={minimum height=1.5em},
   flow/.style={line width=##1*\flowwidth},
   blocklbl/.style={rotate=90,font=\bf,inner ysep=1.5ex},
  ]

  \node[draw,minimum width=5em, rounded corners,microstate](particlesIn) {particles $\mathcal{D}$};
  \node[trapezium, trapezium angle=60, minimum width=5em, below=2\distance of particlesIn, draw, minimum height=4em, trapezium stretches=true] (conv1x1) {$1\times1$ conv};
  \node[draw,minimum width=7em, below=of conv1x1](pool) {max pool};
  \node[draw,minimum width=5em, below left = \distance and -2em of pool](meanMLP) {MLP};
  \node[draw,minimum width=5em, below right = \distance and -2em of pool](varMLP) {MLP};
  \node[draw,minimum width=5em, below=1.5em of varMLP, rounded corners](var) {std dev. $\sigma$};
  \node[draw,minimum width=5em, below=1.5em of meanMLP, rounded corners](mu) {mean $\mu$};
  \node (aboveZ) at ($(var)!0.5!(mu)$) {};
  \node[draw,minimum width=5em, below=2\distance of aboveZ, rounded corners,latent](z) {latent $z$};
  
  \node[draw,minimum width=5em, above right = 1em and 7.5em of z](decMLP) {MLP};
  \node[trapezium, trapezium angle=-60, minimum width=5em, above=of decMLP, draw, minimum height=4em, trapezium stretches=true] (deconv) {deconv 3D};
  \node[draw,minimum width=5em, above=2\distance of deconv, rounded
  corners,microstateApprox](particlesOut) {particles $\mathcal{D}'$};
  
  \node[draw,minimum width=5em, above left= 1em and 7.5em of z, minimum height=7em](inn) {INN};
  \node[draw,minimum width=4em, above left= 2\distance and -4em of inn, , rounded corners, physicalRed](rad) {radiation $I$};
  \node[draw,minimum width=1.5em, above right= 2\distance and -.0em of inn, gray, rounded corners](normal) {$\mathcal{N}$};

  \draw[flow=20] (particlesIn) -- (conv1x1);
  \draw[flow=25] (conv1x1) -- (pool);
  \draw[flow=6] (pool) -- (meanMLP);
  \draw[flow=6] (pool) -- (varMLP);
  \draw[flow=4] (meanMLP) -- (mu);
  \draw[flow=4] (varMLP) -- (var);
  \draw[flow=4] (mu) -- (z);
  \draw[flow=4] (var) -- (z);
  \draw[flow=4] (z.east) to[bend right] (decMLP.south);
  \draw[flow=4] (z.west) to[bend left] (inn.south);
  \draw[flow=3] (inn) -- (rad);
  \draw[flow=1] (inn) -- (normal);
  \draw[flow=6] (decMLP) -- (deconv);
  \draw[flow=10] (deconv) -- (particlesOut);

  \node[draw,encoder,fit=(conv1x1)(meanMLP)(varMLP)(var)(mu),inner sep=1em] (encoder) {};
  \node[draw,decoder,fit=(deconv)(decMLP),inner sep=1em] (decoder) {};
  \node[draw,inverter,fit=(inn),inner sep=1em] (inner) {};

  \node[blocklbl,font=\color{encoder}\bf\large,anchor=north east,xshift=-1em]
   at (encoder.north west) {encoder};
  \node[blocklbl,font=\color{decoder}\bf\large,anchor=north] at (decoder.east) {decoder};
  \node[blocklbl,font=\color{inverter}\bf\large,anchor=south] at (inner.west) {inversion};
  
 \end{tikzpicture}
 }
}

%% file: hsl.tex
\newcommand\FPeval[2]{
 \edef#1{\fpeval{#2}}
}
\newcommand\fpmod[2]{
 #1 - floor(#1 / #2, 0) * #2
}
\newcommand\definecolorHsl[4]{
 \FPeval{\C}{(1 - abs(2* #4 - 1)) * #3}
 \FPeval{\Hprime}{#2 / 60}
 \FPeval{\X}{\C * ( 1 - abs(\fpmod{\Hprime}{2} - 1))}
 \FPeval{\R}{0}
 \FPeval{\G}{0}
 \FPeval{\B}{0}
 \if1\fpeval{\Hprime < 1}
  \FPeval{\R}{\C}
  \FPeval{\G}{\X}
 \else
  \if1\fpeval{\Hprime < 2}
   \FPeval{\R}{\X}
   \FPeval{\G}{\C}
  \else
   \if1\fpeval{\Hprime < 3}
    \FPeval{\G}{\C}
    \FPeval{\B}{\X}
   \else
    \if1\fpeval{\Hprime < 4}
     \FPeval{\G}{\X}
     \FPeval{\B}{\C}
    \else
     \if1\fpeval{\Hprime < 5}
      \FPeval{\R}{\X}
      \FPeval{\B}{\C}
     \else
      \FPeval{\R}{\C}
      \FPeval{\B}{\X}
     \fi
    \fi
   \fi
  \fi
 \fi
 \FPeval{\m}{#4 - \C/2}
 \FPeval{\R}{\R+\m}
 \FPeval{\G}{\G+\m}
 \FPeval{\B}{\B+\m}
 \definecolor{#1}{rgb}{\R,\G,\B}
}